\documentclass[12pt,preprint]{aastex}

\shorttitle{Carbon in the Draco Dwarf Spheroidal Galaxy}
\shortauthors{Shetrone et al.}

\begin{document}

\title{Carbon Abundances for Red Giants in the Draco Dwarf Spheroidal 
Galaxy\footnote{The data presented herein were obtained at the W.M. Keck 
Observatory, which is operated as a scientific partnership among the California
Institute of Technology, the University of California and the National 
Aeronautics and Space Administration. The Observatory was made possible by the 
generous financial support of the W.M. Keck Foundation.}}

\author{Matthew D. Shetrone}
\affil{McDonald Observatory, The University of Texas at Austin, 
 1 University Station, C1400, Austin, TX 78712-0259, USA}
\email{shetrone@astro.as.utexas.edu}
\author{Graeme H. Smith}
\affil{University of California Observatories\/Lick Observatory, Department 
 of Astronomy \& Astrophysics, UC Santa Cruz, 1156 High St., Santa Cruz, 
CA 95064, USA}
\email{graeme@ucolick.org}
\author{Laura M. Stanford }
\affil{McDonald Observatory, The University of Texas at Austin, 
 1 University Station, C1400, Austin, TX 78712-0259, USA}
\author{Michael H. Siegel}
\affil{Department of Astronomy and Astrophysics, Pennsylvania State University,
  525 Davey Laboratory, State College, PA 16801, USA}
\email{siegel@astro.psu.edu}
\author{Howard E. Bond}
\affil{9615 Labrador Ln., Cockeysville, MD 21030, USA}
\email{bond@stsci.edu}

\begin{abstract}
Measurements of [C/Fe], [Ca/H], and [Fe/H] have been derived from Keck I LRISb
spectra of 35 giants in the Draco dwarf spheroidal galaxy. The iron abundances
are derived by a spectrum synthesis modeling of the wavelength region from
4850 to 5375 \AA, while calcium and carbon abundances are obtained by fitting
the \ion{Ca}{2} H and K lines and the CH G band respectively. A range in
metallicity of $-2.9 \leq {\rm [Fe/H]} \leq -1.6$ is found within the giants
sampled, with a good correlation between [Fe/H] and [Ca/H]. The great majority
of stars in the sample would be classified as having weak absorption in the
$\lambda$3883 CN band, with only a small scatter in band strengths at a given
luminosity on the red giant branch. In this sense the behavior of CN among
the Draco giants is consistent with the predominantly weak CN bands found among
red giants in globular clusters of metallicity ${\rm [Fe/H]} < -1.8$. Over
half of the giants in the Draco sample have ${\rm [Fe/H]} > -2.25$, and among 
these there is a trend for the [C/Fe] abundance to decrease with
increasing luminosity on the red giant branch. This is a phenomenon that
is also seen among both field and globular cluster giants of the Galactic 
halo, where it has been interpreted as a consequence of deep mixing of
material between the base of the convective envelope and the outer limits of
the hydrogen-burning shell.  However, among the six Draco giants observed that 
turn out to have metallicities $-2.65 < {\rm [Fe/H]} < -2.25$ there is no such 
trend seen in the carbon abundance. This may be due to small sample
statistics or primordial inhomogeneities in carbon abundance among 
the most metal-poor Draco stars. We identify a potential carbon-rich
extremely metal-poor star in our sample.  This candidate will require 
follow up observations for confirmation.

\end{abstract}


\section{Introduction}

The Draco dwarf spheroidal (dSph) galaxy is chemically inhomogeneous. A 
spread in heavy element abundance was first revealed by the spectrophotometric 
study of Zinn (1978) and confirmed by the low-resolution spectroscopy of 
Kinman et al. (1980), Stetson (1984), and Smith (1984). The inhomogeneities 
are present across a range of elements including Ca (Lehnert et al. 1992; 
Winnick 2003; Smith et al. 2006), other $\alpha$ elements, and the iron-peak 
elements (Shetrone et al. 1998a, 2001a; Cohen \& Huang 2009; Kirby et al. 
2010). Draco has an integrated visual luminosity of 
$L_V = 2.6 \times 10^5 L_{\odot}$ (Mateo 1998), such that its luminosity and 
stellar mass are comparable to those of a medium-mass globular cluster (Hodge 1964).
The general element spread within Draco and other dSphs has been
interpreted as the result of internal chemical evolution (Zinn 1978; Ikuta \& 
Arimoto 2002; Winnick 2003; Marcolini et al. 2006; Cohen \& Huang 2009; Kirby 
et al. 2011a, 2011b). The presence of dark matter halos in dSphs such as Draco 
(Pryor \& Kormendy 1990; Armandroff et al. 1995; Kleyna et al. 2001, 2002; 
Mashchenko et al. 2006) can help account for why systems of such low stellar
mass can have sustained a prolonged episode of element buildup 
by retaining ejecta from evolving stars and/or by capturing new gas.  Dwarf 
spheroidals are now playing a major role in the context of the hierarchical 
formation and evolution of galaxies, and their internal abundance patterns are 
providing insights into how such evolution took place (e.g., Shetrone et al. 
2001a; Geisler et al. 2005; Robertson et al. 2005; Kirby et al. 2008; 
Frebel et al. 2010).

A contrast between the Draco dSph and globular clusters of similar luminous
mass is striking because clusters of this mass are generally very homogeneous
in many of the $\alpha$ and Fe-peak elements that are inhomogeneous in
Draco. The CNO group elements are a notable exception, since inhomogeneities
in the isotopes of this element group are commonplace within globular clusters,
even in clusters of lower stellar mass than the Draco dSph. Concerning carbon 
and nitrogen certain distinctive patterns have been discerned within globular 
clusters. Stars of similar effective temperature and luminosity within the
same cluster can exhibit very different strengths of the $\lambda$3883 or
$\lambda$4215 CN band in the spectrum (e.g., Norris \& Smith 1981).
The fact that such differences occur among main sequence stars within a 
cluster, as first found by Hesser (1978), suggests that they date from very
early times in cluster history, perhaps originating from a period of 
cluster chemical evolution instigated by stars more massive than the present
main sequence turnoff stars. Among stars on the upper part of the red giant
branch (RGB) there is an apparently separate phenomenon that is evinced as a
decline in mean surface carbon abundance with increasing stellar luminosity 
(e.g., Kraft 1984). The inference is that some interior mixing process is
at work within such stars to transport material from the vicinity of the
hydrogen-burning shell to the base of the convective envelope, whereupon it
can be rapidly convected to the stellar surface.  
See Gratton et al. (2012) for a more extensive review of this subject.

There is much less known about the CN and CH distributions within dwarf
spheroidal galaxies than within globular clusters.   Norris et al. (2010)
have a moderate number of stars with carbon abundances in the Bootes and 
Segue I systems but the stars in these ultra faint dwarf galaxies are 
extremely metal-poor and may not serve as a guide for what the carbon
evolution may look like in a more massive "classical" dwarf galaxy.
If the CN inhomogeneities 
within clusters are of a primordial origin, then it is of interest to know 
whether they are present within the different environments of dSph systems.
Dwarf spheroidals are known to have both metallicity spread and a spread in 
ages (see
the recent review by Tolstoy et al. 2009).   The spread in metallicity
may mean that the most metal-poor stars may have a different primordial 
origin for the C and N compared to the more metal-rich stars.   In addition,
some observations and models, e.g. Revaz et al. (2009), suggest that in some 
dwarf galaxies there may be a spread of ages at a single metallicity. As a
final complication, there are carbon depletions 
of luminous cluster red giants that are the product
of a stellar interior mixing process that appears to also function within
halo field red giants of the Milky Way. If the mixing process is a fundamental
attribute of the evolution of low-mass metal-poor stars then it would also be 
expected to occur among red giants in the dwarf spheroidal satellites of 
the Galaxy. With such questions in mind, the present paper reports upon an
observational study of CN and CH bands in the spectra of red giants in the
Draco dwarf spheroidal system.  

In the Galactic halo a very significant fraction of extremely metal-poor stars 
is carbon enhanced, recently measured to be as high as 32\% by Yong et al. 
(2012) for stars with ${\rm [Fe/H]} < -3$. In general the fraction is supposed
to decrease with increasing metallicities, but several studies still find
high fractions for stars just outside the extremely-low metallicity regime
(25\% for ${\rm [Fe/H]} < -2.5$ by Marsteller et al. 2005, at least 21\% for
${\rm [Fe/H]} < -2.0$ by Lucatello et al. 2006, and 14\% for 
${\rm [Fe/H]} < -2.0$ by Cohen et al. 2005). In recent work by Starkenburg 
et al. (2011) carbon measurements were compiled for nine stars with 
${\rm [Fe/H]} < -2.5$ in the Sculptor dSph (seven of which have 
${\rm [Fe/H]} < -3.0$), and none of these stars was found to be carbon 
enhanced. Based on this sample of nine stars, Starkenburg et al. estimated
that there is a low probability of $\sim$ 2\% -- 13\% that the lack of carbon
enhanced stars in Sculptor is entirely a chance effect. Thus carbon may
vary from star to star at a given metallicity for several reasons 
within dSphs such as Draco, and determining the general trends and the 
reasons why the exceptions stand out requires a large sample.

\section{Observations and Reductions}

Spectra of 35 red giants in the Draco dwarf spheroidal were obtained
with the blue channel of the LRIS spectrograph (Oke et al. 1995; McCarthy 
et al. 1998) on the Keck I telescope. This blue channel is denoted as LRISb 
throughout this paper. Observations were made using the multi-object mode of 
LRISb in which the single long-slit assembly is replaced by a slit mask. The 
results reported in this paper are based on observations acquired on UT 2007 
June 19 of two different slit mask fields. Each slit mask field was exposed on 
for a total of five 1800 s integrations. Since LRIS is a dual-channel 
spectrometer a mirror was used in place of a dichroic to reflect all light to 
the LRISb side. The dispersive element employed for the spectroscopy was the 
400/3400 grism, and the detector was a $2 \times 2{\rm K} \times 4{\rm K}$ 
Marconi E2V CCD with 15 $\mu$m pixels. Sky conditions were clear.  As determined
by the emission lamp lines the delivered resolution was 630.   Because
we used slit masks the wavelength coverage varied depending upon the slit position 
within the field but for those slits in the middle of the mask produced spectra
covering the entire optical range.

Basic astrometric and photometric properties of the stars observed are given in
Table 1. Stars can be identified via the right ascensions and declinations 
listed. Column 1 of the table gives the identification assigned here, whereas 
Column 2 lists alternative designations from Baade \& Swope (1961), Stetson 
(1984) and Winnick (2003). Infrared photometry in Table 1 is obtained from the 
2MASS catalog\footnote{This publication makes use of data products from the 
Two Micron All Sky Survey which is a joint project of the University of 
Massachusetts and the Infrared Processing and Analysis Center/California 
Institute of Technology, funded by the National Aeronautics and Space 
Administration and the National Science Foundation. The database can be found 
at http://www.ipac.caltech.edu/2mass/releases/allsky.} for the $J$, $H$ and 
$K_s$ bands. Metallicities derived by Winnick (2003) are listed in the Column 
headed ${\rm [Fe/H]}_{\rm CaT}$ in Table 1.

Optical photometry for members of Draco in our program is based on CCD frames 
obtained by H.~E. Bond with the Kitt Peak National Observatory 0.9 m telescope 
on 1996 September 22. The T2KA chip at the Ritchey-Chretien focus provides a 
$23^{\prime} \times 23^{\prime}$ field, which encloses the core of the galaxy 
($r_c = 9^{\prime}$; Irwin \& Hatzidimitriou 1995). A combination of $BVI$ 
filters was used with exposure times of  
300, 180, and 300 s, respectively, under photometric conditions. The photometry
was calibrated to the network of $BVI$ standard stars established by 
Siegel \& Bond (2005). Data were reduced using the IRAF CCDPROC pipeline and 
photometry was measured with DAOPHOT/ALLSTAR (Stetson 1987, 1994). DAOGROW 
(Stetson 1990) was used to perform curve-of-growth fitting for aperture 
correction on both program and standard stars. The raw photometry was 
calibrated using the iterative matrix inversion technique described in 
Harris et al. (1981) and Siegel et al. (2002) to translate the photometry 
to the standard system of Landolt (1992) and Siegel \& Bond (2005). The $V$ 
magnitudes plus $(B-V)$ and $(V-I)$ colors derived from this photometry for 
the Draco program stars are listed in Table 1. 

The color-magnitude diagram (CMD) of the observed Draco sample is plotted in 
Figure 1, based on the $(V, B-V)$ photometry from Table 1.
A linear least-squares (lsq) fit to the sequence defined by this sample has
the equation $V_{\rm lsq} = 21.655 - 3.402 (B-V)$, and is shown by the solid 
line in the figure. The standard deviation in the slope of this relation is
0.291. Each star presents a color residual
$\Delta (B-V) = (B-V) - (B-V)_{\rm lsq}$ with respect to this lsq fit.  
In a chemically inhomogeneous system these color residuals would be expected
to correlate with stellar metallicity, an expectation that is testable with
the data.

The spectra were reduced using {\sc iraf} 2dspec routines. To set the final
velocity/wavelength scale the blended stellar lines in the spectra themselves 
were used. The position of each line was determined (using splot in 
IRAF\footnote{IRAF is distributed by the National Optical Astronomy 
Observatory, which is operated by the Association of Universities for Research 
in Astronomy, Inc., under cooperative agreement with the National Science 
Foundation.}) from a synthetic spectrum convolved to the resolution of LRISb
with stellar parameters typical of our sample stars.  
Although flux standard stars were observed with the LRIS 
long slit, the spectra cover a different wavelength range than some of the 
observed spectra, and thus were of limited use for fluxing the 
data. Fluxing was done after the final abundance analysis was completed 
and is discussed in the following section.

\section{Abundance Analysis and Indices}

\subsection{Stellar Parameters} 

For this analysis we used photometric surface gravities and effective 
temperatures and metallicities from the red portion of the spectrum (see 
Section 3.2 for this procedure).  
As an initial guess we used the metallicities
from Winnick (2003) for the 25 stars in common. Winnick (2003) used a 
metallicity calibration from the calcium triplet lines to obtain metallicities 
for the Draco dSph giants. For those stars without a metallicity estimate we 
began with a metallicity of ${\rm [M/H]} = -2$. When an initial estimate of 
the metallicity was more than 0.15 dex different from the derived low 
resolution metallicity another iteration was made with a second determination 
of effective temperature and surface gravity.   

Effective temperatures were obtained for stars in the sample by using 
metallicities, photometry and the calibrations of Ram\'{i}rez \& Mel\'{e}ndez 
(2005) and a reddening of $E(B-V)=0.03$.
The values of $T_{\rm eff}$ listed in Column 8 of Table 2 are an 
average of the temperatures obtained using the $(V-I)$, $(B-V)$, $(V-J)$, 
$(V-H)$, and $(V-K)$ calibrations, after potentially iterating if the initial 
guess of the metallicity was not sufficiently accurate.   

The surface gravity was obtained using the standard equation:
\begin{equation}
{\rm log}\,g = {\rm log}\,g_{\odot} + {\rm log}(M/M_{\odot}) + 0.4(M_{\rm{bol}}
 - M_{\rm{bol}\odot}) + 4\,{\rm log} (T_{\rm{eff}}/T_{{\rm eff}\odot}),
\end{equation}
where the bolometric magnitude is given by
\begin{equation}
M_{\rm bol} = V - (m-M)_V + BC_V.
\end{equation}
The distance modulus, $(m-M)_V$, was adopted as 19.65, 
$M_{{\rm bol}\odot} = 4.75$ , $\log g_{\odot} = 4.44$ cgs, 
$T_{{\rm eff}\odot} = 5790K$. The bolometric correction $BC_V(T_{\rm eff}$) 
was taken from Alonso et al. (1999).

The microturbulence was determined from a calibration of the Lick-Texas 
group (e.g., Kraft \& Ivans 2003) analyses of globular cluster giants and the 
DART (Tolstoy et al. 2006) analyses of dSph giants as follows:
\begin{equation}
v_t = -0.41 \log g + 2.15,
\end{equation}
where $v_t$ is in km s$^{-1}$ and $g$ is in cgs.
Values of surface gravity and microturbulence velocity adopted for each Draco
star to be analyzed are listed in Columns 10 and 11 respectively of Table 2.

\subsection{Analysis Procedure}

The line list used for this project includes atomic, C$_2$, CN, SiH, and MgH 
lines from the Kurucz compilation 
(http://kurucz.harvard.edu/LINELISTS/GFHYPERALL). The CH line list came from 
B.~Plez 2010 (private communication). Model atmospheres were computed without
convective overshooting from the Kurucz (1993) grid, using interpolation 
software developed by A.~McWilliam 2009 (private communication).  

Synthetic spectra were computed using the 2010 scattering version of {\sc MOOG}
(Sneden 1973, Sobeck et al. 2011). The standard version of MOOG treats 
continuum scattering ($\sigma_\nu$) as if it were absorption ($\kappa_\nu$) 
in the source function, i.e., $S_\nu = B_\nu$ (the Planck function), an 
approximation that is only valid at long wavelengths. At shorter wavelengths, 
Cayrel et al. (2004) have shown that the scattering term must be taken into 
account such that the source function becomes 
$S_\nu = (\kappa_\nu B_\nu +\sigma_\nu J_\nu)/(\kappa_\nu +\sigma_\nu)$.
For metal-poor red giants, the difference in iron line abundances usually 
becomes notable ($> 0.05$ dex) below 5000 \AA.  The scattering corrections are 
negligible for red lines ($>5000$ \AA), but can approach 0.4 dex for resonance
lines in the blue.

Some assumptions about element abundance ratios were needed for the analysis.
For oxygen and the other alpha elements the abundances were assumed to be 
[$\alpha$/Fe]=[Ca/Fe], although the choice of [O/Fe] had little impact on 
the final carbon determination (a change of [O/Fe] of 0.3 dex from that 
assumed by setting [O/Fe] = [Ca/Fe] causes the derived [C/Fe] to change 
by 0.06 dex).  The carbon isotope ratio was assumed to be 
${\rm C}^{12}/{\rm C}^{13} = 10$. Little is known regarding the isotopic 
carbon ratio among red giants in dwarf galaxies so we adopt a value between
that found among the brightest halo red giants and those near the bump in the 
luminosity function, see Gratton et al. (2000) for field stars and Smith
et al. (2007), Shetrone (2003) and Recio-Blanco \& de Laverny (2007) for
globular cluster stars.

The iron abundance was determined from the 4850-5375 \AA\ region by using
a spectrum synthesis approach. Synthetic spectra with different iron abundances
were compared to an observed spectrum, excluding regions with bad sky 
subtraction or cosmic rays. The abundance is taken from the synthetic spectrum 
that produced the smallest residuals about a constant value that was allowed 
to vary from comparison to comparison to compensate for errors in the 
continuum normalization.

The [Ca/Fe] abundance was determined from the \ion{Ca}{2} H and K lines via a
technique similar to that used to determine the iron abundance except that 
the [Ca/H] abundance was now allowed to vary instead of [Fe/H]. 
Three pixels at the core of both the 
H and K lines were excluded from the comparisons since those regions were 
found to be not well fit by the model atmospheres and techniques used.     

After [Fe/H] and [Ca/H] were determined the G band at $\sim 4300$ \AA\ was 
modeled to derive carbon abundances using the same techniques that
were applied to the 4850-5375 \AA\ region for iron abundance measurement 
except that only carbon was allowed to vary. Figure 2 shows a small section 
of the observed spectrum for star 161, displayed as the black points, 
along with several synthetic spectra with different [C/Fe] ratios.  

The [Fe/H], [Ca/H] and [C/Fe] abundances derived from the synthetic spectra
analyses of the LRISb spectra are listed in Columns 4, 6 and 8 of Table 2.
A low order polynomial fit of the initial spectrum divided by the final 
synthetic spectrum was divided back into the initial spectrum to produce
the final normalized observed spectrum.  

\subsection{Error Analysis}

Four stars, denoted 361, 409, 410, 427 in Table 1, were observed through both 
of the slit masks employed. These multiple observations allow us to compare
the abundance results for internal consistency. The average of differences
in [Fe/H], [Ca/H], and [C/Fe] between these observed pairs is 0.04, 0.05, and 
0.10 dex, respectively.   Our estimate for the typical fitting 
(measurement) errors for these three abundances 
are 0.08, 0.06, and 0.05 dex, respectively.   

The formal internal error of the mean of the effective temperatures is 
small. In order to look for systematic and external errors we compared the 
temperatures derived in this work with spectroscopic temperatures derived in
the high resolution analyses of Shetrone et al. (2001a) and Cohen \& Huang
(2009) for five stars that we have in common. Our temperatures are 
33 $\pm$ 56 K cooler than the literatures values with a standard deviation
of 110 K. This standard deviation is far larger than the internal errors on 
the photometric effective temperature for these stars. Thus we inflate our 
internal scatter error by 80 K (combined in quadrature) to cover the possible 
external errors. These modified temperature errors $\epsilon(T_{\rm eff})$ 
are listed in Column 12 of Table 2.

The uncertainty on the abundances were determined using the standard
practice of propagating the uncertainties in stellar parameters and
combining in quadrature. The temperature was varied by the amount
listed in Column 12 of Table 2. The gravity and microturbulence
were varied by 0.2 dex and 0.3 km s$^{-1}$ respectively. Since [Fe/H] is
relatively insensitive to mistakes in the model atmosphere metallicity
the error in the model metallicity can be formulated by combining the
error in gravity, temperature and microturbulence for [Fe/H]. By contrast,
[Ca/H] and [C/Fe] are more sensitive to errors in the model atmosphere
metallicity thus we also include the error in [Fe/H] along with the
errors in gravity, microturbulence and effective temperature in the
quadrature sum. With each new model the features were refit and new 
abundances determined. This analysis was performed for several stars
spanning the range in temperature and metallicity of the Draco sample, 
resulting in an average error formula for [C/Fe] of
$\epsilon{\rm [C/Fe]} = [0.0194 + (\epsilon{\rm [Fe/H]})^{2} 
       + (0.0005 \times \epsilon(T_{\rm eff}))^{2}]^{1/2}$
and
$\epsilon{\rm [Ca/H]} = [0.0013 + (0.4 \times \epsilon{\rm [Fe/H]})^{2} 
          + (0.0006 \times \epsilon(T_{\rm eff})^{2}]^{1/2}$
for [Ca/H]. The uncertainties thereby computed for the [Fe/H], [Ca/H] and
[C/Fe] abundances are listed in Columns 5, 7 and 9 respectively of Table 2.

\subsection{Comparisons with Other Abundances}

The veracity of the [Fe/H] abundances from Column 4 of Table 2 can be tested
via several comparisons. The first comparison is shown in Figure 3, in which 
[Ca/H] as derived from the \ion{Ca}{2} H and K lines is plotted versus [Fe/H] 
from \ion{Fe}{1} lines. There is a clear correlation with a slight offset in 
the sense that [Ca/H] is systematically $\sim0.05$ dex greater than [Fe/H]
except at the highest metallicities in the sample with ${\rm [Fe/H]} > -1.8$. 
Various symbols are used in Figure 3 to denote three different metallicity 
ranges based on the LRISb [Fe/H] abundances: filled squares 
(${\rm [Fe/H]} < -2.25$), 
filled circles ($-2.25 \leq {\rm [Fe/H]} \leq -1.85$), and filled triangles 
(${\rm [Fe/H]} > -1.85$).  These symbols for stars of different metallicity 
are also used in other figures of this paper. 

Winnick (2003) derived [Fe/H] for a large sample of Draco giants from 
calibrating the strength of the \ion{Ca}{2} near-infrared triplet lines. There 
are 25 stars in our Draco program for which Winnick derived such abundances, 
that are denoted here as ${\rm [Fe/H]}_{\rm CaT}$ (Table 1). A plot of 
${\rm [Fe/H]}_{\rm CaT}$ versus the [Fe/H] abundances from Column 4 of Table 2
is presented in Figure 4. By and large the comparison is quite favorable, with 
the data scattering about the line corresponding to equality. The LRISb 
metallicities are identical with ${\rm [Fe/H]}_{\rm CaT}$ to within the 
0.03 dex error on the mean difference (the standard deviation being 0.17 dex). 
There is perhaps a tendency for the Ca triplet values of [Fe/H] (as calibrated 
by Winnick) to be systematically less than the LRISb abundances by $\sim 0.1$ 
dex for the lowest-metallicity stars in the sample.   This is not entirely 
surprising as more recent non-linear calibrations for the Ca triplet have
been suggested for very metal-poor stars, see Starkenburg et al. (2010).

Iron abundances have been measured from medium-resolution spectra for several
of the stars in Table 2 by Kirby et al. (2010). The comparison between the 
Kirby et al. [Fe/H] values and our sample is given in Table 3 and
shown in Figure 4 as blue triangles. The general 
agreement is very good ($+0.04 \pm 0.05$, $\sigma = 0.16$). A comparison can 
also be made based on a small overlap with high-resolution studies by Shetrone 
et al. (2001a), Fulbright et al. (2004), and Cohen \& Huang (2009). In the case
of the three giants with high-resolution abundances of ${\rm [Fe/H]} < -2.0$ 
dex the agreement with the LRISb-based [Fe/H] values is excellent 
($-0.05 \pm 0.06$ dex). However, for the two more metal rich giants (187 and 
240) the agreement is quite poor, with the LRISb [Fe/H] values being 
$0.39 \pm 0.01$ less than those of Shetrone et al. (2001a). The cause of 
this larger discrepancy is not known.    The agreement between our 
[Ca/Fe] abundance ratios and those in the two high resolution samples 
is excellent ($-0.04 \pm 0.06$).  

Another check upon the [Fe/H] abundances from Table 2 is to see whether they
correlate with the position of stars on the RGB of the 
CMD of Draco. The color residual $\Delta (B-V)$
defined in Section 2 is plotted in Figure 5 versus the LRISb [Fe/H] abundance
from Column 4 of Table 2. A correlation is evident, albeit with notable
scatter. One interpretation of Figure 5 might be that the relation between
metallicity and color on the RGB of Draco is not 1-to-1,
particularly among the lowest metallicity giants. This complication was
first discussed by Stetson (1980) and Zinn (1980). In a system such as Draco
in which there is a metallicity spread it can be hard to distinguish between 
RGB and asymptotic giant branch (AGB) stars, making it more difficult 
to interpret any relation between color and metallicity.  Scatter in Figure 5 
can also result from another source, namely that the linear fit in Figure 1 
does not properly represent the curvature of the RGB in the 
CMD. This could be particularly important for stars at the 
fainter end of our sample where the locus of the RGB is becoming 
steeper than at higher luminosities. Other potential sources of confusion 
between color and metallicity are spreads in age and a spread in 
the [$\alpha$/Fe] abundance ratio (both as a function of metallicity
and at a fixed metallicity).

There are very few studies of [C/Fe] in dwarf galaxies. Cohen \& Huang (2009) 
have a sample of eight stars, one of which we have also observed. That 
star in common is star 161, called 3157 in Cohen \& Huang (2009) and shown
in Figure 2. We determine a [C/Fe] of $-0.55$ while Cohen \& Huang (2009)
derive $-0.29$.  The solar Fe and C abundances adopted by 
Cohen \& Huang (2009) are 7.45 and 8.59, respectively, while we use
the default Fe and C abundance found in MOOG, i.e., 7.52 and 8.56. When 
this is accounted for
the discrepancy is reduced slightly, from 0.26 to 0.22 dex. As a further 
test we used the
Cohen \& Huang (2009) stellar parameters for star 161 and recovered their 
preferred [C/Fe] abundance to within our measurement error of 0.05 dex. This
shows that our technique and spectra can derive abundances having an accuracy
dominated by modeling and systematic errors.

\section{Indices}

To flux calibrate the data the spectral flux continuum output option from 
MOOG 2010 (Sneden 1973) was used as given for the final stellar model and 
abundances from our analysis. With this option the flux values used within
MOOG to normalize the synthetic spectra were sent to the output. We multiplied 
the final normalized observed spectrum by these synthetic flux values to 
produce a fluxed spectrum. 

Two indices were measured from the fluxed LRISb spectra. Denoted $S(3839)$ and 
$S_2$(CH) they quantify the strength of the $\lambda$3883 CN and $\lambda$4300
CH band respectively. The definitions of these indices are
\begin{equation}
S(3839) = -2.5\, \log \, \int_{3846}^{3883} I_{\lambda} \, d\lambda \, / 
  \int_{3883}^{3916} I_{\lambda} \, d\lambda
\end{equation}
from Norris et al. (1981), and 
\begin{equation}
S_2{\rm (CH)} = -2.5\, \log \,\frac{ \int_{4280}^{4320} I_{\lambda} \, d\lambda
  \,}{ \frac{1}{2} (\int_{4050}^{4100} I_{\lambda} \, 
 d\lambda + \int_{4320}^{4360} I_{\lambda} \, d\lambda)}
\end{equation}
from Martell et al. (2008b). Values of these indices for the Draco stars 
observed with LRISb are listed in Table 2.

Two sets of index values were obtained for the stars 361, 409, 410, 427, since 
they were observed through both of the slit masks employed. The agreement among
the two sets of index measurements is good, such that the mean index values 
for these stars are listed in Table 2. The greatest difference in $S(3839)$ 
is 0.045 for star 410, while for the other three stars it is less than 0.015. 
In the case of the $S_2$(CH) index the greatest difference is 0.029 for star 
409, whereas for the other three stars it is less than 0.010. 

\section{Results}

\subsection{The [Fe/H] Abundance}

The iron abundances from this work show a range among the Draco stars observed
of $-2.9 \leq {\rm [Fe/H]} \leq -1.6$. This is very similar, but slightly less,
than the range from $-3.0$ to $-1.5$ dex found by Cohen \& Huang (2009) and 
Shetrone et al. (1998a), and $-2.97$ to $-1.44$ found by Shetrone et al. 
(2001a), all from HIRES spectroscopy for smaller samples of red giants. It is 
not as broad as the ranges from $-3.1$ to $-1.0$ dex found by Winnick (2003) 
and Kirby et al. (2011b) on the basis of WIYN Ca triplet line spectroscopy and 
Keck DEIMOS spectroscopy respectively. Using Str\"{o}mgren photometry of red 
giants Faria et al. (2007) derived a metallicity distribution function for 
Draco that extends mainly from [Fe/H] = $-1.4$ to $-2.2$, with a mean of 
$-1.74$ dex, and a small fraction of stars outside this range. Our LRISb 
abundance range is either weighted to more metal-poor stars than the 
Faria et al. (2007) distribution, or else there is an offset between the 
abundance scale of Faria et al. (2007) and that of the present work.  The
referee pointed out that Faria et al. used the older calibrations of 
Hilker (2000) and Anthony-Twarog \& Twarog (1994), which do not cover the 
metal poor tail. Re-calibrations of the Draco data to newer 
Stromgren-metallicity scales (Aden et al. 2009; Calamida et al. 2007, 2009) 
could alleviate the present discrepancy. 

The highest metallicity stars in our Draco sample have [Fe/H] similar to that 
of the globular clusters M3 and M13, and close to the peak in the metallicity
distribution of globular clusters of the Galactic halo. The most metal-poor 
globular clusters in the Milky Way, such as M15 and M30, have 
${\rm [Fe/H]} \sim -2.4$ (Sneden et al. 1997; Carretta et al. 2009). Stars of 
lower metallicity than this are present in the LRISb Draco sample. Hence the 
stars in the Draco dSph evince a metallicity spread that overlaps substantially
with the metallicity distribution of halo globular clusters. A comparison
between the color spread on the RGB of the Draco system and the 
range among halo globular clusters is consistent with this conclusion (Zinn 
1980; Aparicio et al. 2001; Bellazzini et al. 2002; Cioni \& Habing 2005).

Star 235 is worth noting. It is alternatively designated Draco 119 or 195-119 
in some papers after the notation used in the color-magnitude study of Baade \&
Swope (1961). The peculiar low-metallicity nature of this giant was first 
evident in the DDO photometry of Hartwick \& McClure (1974) and the 
low-resolution spectroscopy of Kinman et al. (1980). It had been included 
in the survey of Zinn (1978) but had not shown an unduly low metallicity from 
his spectrophotometry, although the [Fe/H] derived by Zinn was 
amongst the lowest in his program. The abundance derived for it here is
close to [Fe/H] = $-3.0$, making it not only the lowest-metallicity star
in our sample, but also of a lower metallicity than any globular cluster. 
High-resolution spectroscopic abundance analyses have shown that this is
indeed the case (Shetrone et al. 1998a, 2001a; Fulbright et al. 2004). From 
a historical perspective Draco 119 is one of the earliest known examples to 
suggest that dwarf spheroidal galaxies contain some stars that are more 
metal-poor than any in the Milky Way globular cluster system.  More recent
and larger samples, e.g. Kirby et al. (2011b), have found more stars 
with [Fe/H]$ < -3.0$ showing that sample selection and size are critical 
in understanding the metal-poor tail of the metallicity distribution function.

\subsection{The CN Index}

The CN index $S(3839)$ is plotted versus $M_{\rm bol}$ in Figure 6. Symbols
are coded according to metallicity as in Figure 3. The CN index on average
increases with increasing luminosity, typical of trends seen in globular 
clusters. The spread in CN index at a given $M_{\rm bol}$ is 0.06 or
less with the exception of only a few stars. There is little to distinguish 
between the stars in the metallicity ranges $-2.65 \leq {\rm [Fe/H]} < -2.25$ 
and $-2.25 \leq {\rm [Fe/H]} \leq -1.85$ within the figure, and most of these 
giants would be considered to have weak CN bands. 

Given that the majority of Draco stars in our LRISb sample have metallicities 
of ${\rm [Fe/H]} < -1.8$, perhaps the most appropriate globular 
clusters with which to compare the Draco result are metal-poor systems such as
M55 (Smith \& Norris 1982; Briley et al. 1993), M53 (Martell et al. 2008a) and 
NGC 5466 (Shetrone et al. 2010), all of which show only limited dispersions in 
$S(3839)$ at a given magnitude on the RGB. Globular clusters with
metallicities of ${\rm [Fe/H]} \geq -1.7$ can show bimodal CN variations with 
spreads of up to 0.4 mag in the $S(3839)$ index among otherwise similar giants.
However among globular clusters more metal-poor than this, and particularly at 
the metallicities of most giants in Figure 6, the CN inhomogeneities within 
globular clusters tend to be much more muted. Thus there is nothing necessarily
unusual about Draco with respect to the behavior of the CN bands of its red 
giants. 

Nonetheless there are some stars (325, 348, and 354) that do seem to show 
stronger $\lambda$3883 CN bands than other Draco giants of similar 
$M_{\rm bol}$. Draco 325 and 354 have values of the $S(3839)$ index that are 
0.08-0.10 mag larger than that of star 276. These three stars have similar 
[Fe/H] abundance ($-1.6$ to $-1.70$ dex), and CN inhomogeneities are 
commonplace within globular clusters of such metallicity. The most CN-rich 
giant in the sample is 348 (${\rm [Fe/H]} = -1.9$), with an $S(3839)$ index 
more than 0.1 mag larger than other Draco giants of similar [Fe/H] and 
luminosity. Star 348 has stronger CN than any of the three giants with 
${\rm [Fe/H]} > -1.8$ in the LRISb sample (compared to which it has a  
similar $M_{\rm bol}$). Thus the CN band of 348 is suggestive of either 
a selective carbon or nitrogen abundance enhancement.  

\subsection{The Carbon Abundance Trends}

Behavior of the $\lambda$4300 CH-band index $S_2$(CH) is shown versus 
bolometric magnitude in Figure 7, the plotted points again being coded with 
the same metallicity-dependent symbols as in Figure 3. First impressions from 
the figure are of a random scatter, however if stars are considered according 
to their [Fe/H] metallicity then a few features show up. In the case of either 
a constant carbon abundance, or a carbon abundance that decreases with stellar 
luminosity, the $S_2$(CH) index will decrease in value with increasing 
luminosity on the upper RGB (Martell et al. 2008b). A trend of 
this sort can be discerned, on average, among the giants with 
$-2.25 < {\rm [Fe/H]} < -1.85$. By contrast, the $S_2$(CH) index increases 
with increasing luminosity among the metal-poorer group of giants with 
$-2.65 < {\rm [Fe/H]} < -2.25$. Whether this possible trend is a general 
property of this particular metallicity subgroup in Draco remains to be seen, 
the trend could arguably be a consequence of small sample statistics and the 
inclusion of one star 621 that has the weakest CH band in our entire
Draco program. Both the CN and CH index increase, on average, with increasing
luminosity for the [Fe/H] = $-2.25$ to $-2.65$ subgroup, however this need not
imply an abundance-driven correlation between the stars within this group 
because the $S(3839)$ index is sensitive to effective temperature. 

There are, however, hints of a CN-CH correlation among at least some of the 
stars in the LRISb Draco sample. The giant with the largest $S(3839)$ index 
(348) also has the largest $S_2$(CH) index. Among the three giants with 
${\rm [Fe/H]} > -1.85$ there is a range of 0.1 mag in $S_2$(CH) such that the
giant with the smallest CH index also has the smallest $S(3839)$. The most 
metal-poor star in the sample with ${\rm [Fe/H]} = -3.0$ has a notably weaker
CH band than other giants of similar bolometric magnitude. 

We have also determined the [C/Fe] abundance ratio using synthetic spectra and 
thus are somewhat independent of the CH indices (although they do come from the
same spectra and the fluxing relies on the synthetic fit from MOOG). These 
abundance ratios are shown in Figure 8 versus stellar bolometric magnitude.
The symbols are again coded according 
to metallicity as in Figure 3. As with Figure 7 a first glance gives the 
impression of a scatter diagram, but when the various metallicity subgroups 
are considered a notable trend is evident. Among the giants with 
${\rm [Fe/H]} > -2.25$ the average [C/Fe] on the RGB declines 
with increasing stellar luminosity. The three giants with 
${\rm [Fe/H]} > -1.8$, as well as the very metal-poor giant 235, follow
much the same trend displayed by giants in the metallicity range
$-2.25 < {\rm [Fe/H]} < -1.85$. Therefore, among the majority of stars in the
LRISb Draco sample there is evidence of a declining carbon abundance with
advancing evolution up the RGB. The notable exception to this
general trend concerns the stars with ${\rm [Fe/H]} < -2.25$, among
which if there is any trend at all it is one of increasing [C/Fe] up the
RGB.

All but one of the Draco stars in our LRISb sample are found to have 
${\rm [C/Fe]} < 0$. This result seems analogous to the depleted carbon 
abundances typically encountered among stars on the upper half of the RGBs
of globular clusters. The range of carbon abundances within the 
bulk of our Draco sample and the result that [C/Fe] is predominantly less than
0.0 dex is consistent with the findings of Cohen \& Huang (2009). We discuss 
the carbon rich stars (${\rm [C/Fe]} > 0.0$) in the next subsection.
Cohen \& Huang (2009) found a trend between 
[C/Fe] and [Fe/H] among the eight stars in their Draco sample (see their 
Figure 3), but such is not the case with the results of the present sample. 
A plot of [C/Fe] versus [Fe/H] from the LRISb data (Figure 9) shows a 
notable spread in [C/Fe] at a given [Fe/H] but no obvious correlation.

There appears to be a dispersion in [C/Fe] of up to 0.6 dex among giants
of similar [Fe/H] and/or similar $M_{\rm bol}$, and some of this dispersion 
may be of a primordial nature, particularly among those giants with
${\rm [Fe/H]} < -2.25$. However, the luminosity-dependent trend seen 
among the other Draco stars in Figure 8 is suggestive that some type of deep 
mixing process is acting within the Draco giants so as to bring CN(O)-processed
material up to their surfaces from the interior hydrogen-burning shell. In
this sense the Draco stars appear to be behaving in a manner analogous to
the red giants of metal-poor globular clusters such as M92, M15, and NGC 5466 
(Carbon et al. 1982; Trefzger et al. 1983; Langer et al. 1986; Bellman et al. 
2001; Shetrone et al. 2010), as well as halo field giants of comparable [Fe/H] 
metallicity (Gratton et al. 2000). 

A direct comparison between the run of [C/Fe] with bolometric magnitude 
$M_{\rm bol}$ for the Draco stars in our sample
with $-2.25 \leq {\rm [Fe/H]} \leq -1.85$ and for red giants in NGC 5466 is
shown in Figure 10. The carbon abundances for NGC 5466 are taken from 
the ``Main sample'' 
listed in Table 2 of Shetrone et al. (2010), with a distance modulus of
$(m-M)_V = 16.15$ being adopted for the cluster. We have chosen to exclude the
single CH star in NGC 5466 to avoid confusion of the main trend.  
NGC 5466 makes for a good comparison because it is similar in 
metallicity to the Draco giants plotted in Figure 10, and because the RGB 
shows only a small dispersion in [C/Fe] at a given absolute 
magnitude. The Draco giants exhibit a similar trend of decreasing [C/Fe] with 
increasing luminosity.  The one notable difference is that there is a larger 
dispersion (rms of 0.19 dex) in [C/Fe] at a given $M_{\rm bol}$ in Draco than 
in NGC 5466. However, given the larger individual errors in Table 2 the reduced
$\chi^2 = 0.86$ is consistent with all of the given scatter being due to 
our modeling and systematic errors. Figure 10 suggests that at a given 
$M_{\rm bol}$ on the RGB many of the Draco stars are comparable in 
[C/Fe] to giants of NGC 5466, although some Draco stars do extend 
to lower carbon abundances than their NGC 5466 counterparts of 
similar $M_{\rm bol}$. The open symbols in Figure 10 are 
taken from the Draco sample of Cohen \& Huang (2009) and the crosses are
Ursa Minor dSph giants from Cohen \& Huang (2010). 
While these samples are small and limited to the brightest giants both 
are consistent with the other Draco and NGC 5466 stars plotted.

Figure 11 is analogous to Figure 10 except that it shows the Draco and Ursa 
Minor stars with ${\rm [Fe/H]} > -1.85$. We have kept the NGC 5466 points in
this figure as a guide to the expected range for the  ``normal'' C-depletion
pattern even though NGC 5466 is more metal-poor than the dwarf spheroidal 
stars plotted. All of the Draco and Ursa Minor points fall among the 
NGC 5466 points. Cohen \& Huang (2009) showed that the most metal-rich Draco 
stars exhibit significant {\it s}-process enhancement suggesting contribution
from AGB stars, while Cohen \& Huang (2010) showed that the most metal-rich
Ursa Minor giants show no significant {\it s}-process enhancement and thus are 
more like typical Type II halo giants. Despite these differences there
does not seem to be any detectable (within our observational uncertainties) 
carbon enrichment associated with this {\it s}-process enhancement
in Draco. This may 
place useful constraints on future chemical enrichment models.

Whether the lack of a carbon-luminosity relation among the giants in our 
sample with ${\rm [Fe/H]} < -2.25$ could be due to small sample 
statistics or the additional effect of primordial inhomogeneities in the carbon
abundance is not known. Given that the carbon-enhanced star 589 in our sample 
has a measured [Fe/H] also in this range, it may be that some of these 
lower-metallicity giants are exhibiting a much more modest carbon-star-like 
phenomenon that has been obscured even further by deep mixing. Perhaps one is 
reminded of the so-called ``insipid CH star'' identified in the Sculptor dSph 
by Shetrone et al. (1998b). The large frequency of these C-rich objects in 
dSphs has been noted before, e.g., Shetrone et al. (2001b), Cohen \& Huang 
(2009). Shetrone et al. (2001b) compared the CH star frequency in dSphs and 
found it to be more than an order of magnitude larger than in globular 
clusters. Cohen \& Huang (2009) point out that most of these CH stars are more 
metal-poor than the most metal-poor globular clusters and thus a better 
comparison would be the metal-poor field halo stars where the frequency of 
very C-rich stars is about 20\%. The very C-rich stars are discussed further 
in Section 5.4.

The three brightest stars in Figure 8 have bolometric magnitudes close to
the theoretical tip of the first-ascent RGB. A comparison can 
be made with theoretical isochrones from the Dartmouth Stellar Evolution 
Database (Web site \texttt{stellar.dartmouth.edu}) described by Dotter et al. 
(2008). The isochrones are derived from stellar evolutionary tracks computed by
the Dartmouth Stellar Evolution Program (Bjork \& Chaboyer 2006; Dotter et al. 
2007). The tip of the RGB of the Dartmouth isochrone for 
$Y=0.2454$, 12 Gyr, [Fe/H] = $-2.00$, and [$\alpha$/Fe] = 0.20 is located at
$M_V = -2.61$, $\log L/L_{\odot} = 3.25$, $M_{\rm bol} = -3.38$. The two
giants with the highest [C/Fe] abundance in the metallicity range 
$-2.65 < {\rm [Fe/H]} < -2.25$ are close in luminosity to this limit, and as 
such may potentially be AGB stars. Their high carbon
abundances might be related to evolutionary effects during the AGB
phase of evolution, such as the dredge-up of carbon from the deep
interior. These giants may also be slightly younger, and of a slightly higher 
mass, than most red giants in our Draco sample. Such differences could 
contribute to why these stars do not follow the pattern of declining carbon 
abundance with advancing luminosity exhibited by other giants in Draco. Such 
stars may also be responsible for the conclusions of Smith et al. (2006) who 
found that the most luminous stars in Draco tend to higher carbon band 
indices than what might be expected based on globular cluster results.  

\subsection{Exceptionally Carbon-rich Stars}

As noted above, there are a few exceptions to the general trend that
[C/Fe] is less than 0.0 dex on the RGB in Draco.   The Cohen \& Huang (2009)
sample contains one red giant with [Fe/H] = $-3.0$ that is enhanced in carbon 
([C/Fe] = 0.3 dex). Star 589 in our LRISb sample, for which we find 
[C/Fe] = 0.60, may be another example of such a star. Not only is the [C/Fe] 
ratio above solar for both of these stars but their [Fe/H] abundances are 
similar as well. Unfortunately, due to the position of star 589 with respect 
to the center of our slit plate we did not get a full spectrum and we have 
no information on the strength of the Ca H and K lines or the $S(3839)$ index.
Carbon stars are known in Draco and have been studied in a number of papers 
(Aaronson et al. 1982; Shetrone et al. 2001b; Smith et al. 2006; Abia 2008).   
Abia (2008) confirm that the very luminous C-rich stars in Draco are metal-poor
(${\rm [M/H]} \le -2$) and at least two of the three Draco stars in their 
survey are 
variables suggesting that they are thermally pulsing AGB carbon stars. In 
contrast, star 589 is far less luminous, however, it does lie blueward of the 
bulk of the Draco RGB population suggesting that it might be an AGB star.  
A comparison of the Abia (2008) C-rich stars to the C-rich star of Cohen \& 
Huang (2009) is not so clear. They have similar surface gravities and optical 
luminosities but vastly different temperatures, and the Abia (2008) objects 
are more carbon enhanced. By the definition of Aoki et al. (2007) all five of 
these C-rich stars could be labeled as carbon-rich extremely metal-poor (CEMP) stars.

The small survey by Starkenburg et al. (2011) of very metal-poor stars
in the Sculptor dSph did not reveal any CEMP stars, however their survey
covered ${\rm [Fe/H]} < -2.5$ which would not have included stars such as the 
insipid CH star in Sculptor found by Shetrone et al. (1998b). 
If the same criterion of 
${\rm [Fe/H]} < -2.5$ was applied to our Draco sample and if followup 
observations
reveal star 589 to be a true CEMP star we would have one
CEMP star out of five stars.  If we combine our sample with that of 
Cohen \& Huang (2009) then we would have one or two CEMP stars out of a sample
of seven Draco stars with [Fe/H] $< -2.5$.  Draco is not the only dwarf galaxy 
to exhibit some CEMP stars,
Norris et al. (2010) and Frebel et al.
(2010) find CEMP stars in several ultra faint dwarf galaxies.  
Why these dwarf galaxies should exhibit CEMP stars while the Sculptor dSph 
does not is unclear.   Shetrone et al. (2001b)
note that the CH stars they find are redward of the majority of stars in the 
CMD due to the Bond-Neff effect (Bond \& Neff 1969), 
where extra opacity sources, presumably strong CN and CH bands in
these metal-poor stars, remove 
light from the blue part of the spectrum. 
Perhaps this selection plays some role in 
systems where the CEMP targets are identified from a CMD. 
Further possible CH stars from the literature (e.g. Shetrone et al. 2001b, 
Smith et al. 2006) could be used to follow up the CEMP fraction in Draco.

\section{Conclusions}

In summary, the CN and CH molecular bands in the spectra of Draco red giants 
behave in ways that are not dissimilar to that of globular cluster giants of 
comparably low metallicity. There is evidence for several CN-enhanced giants 
but the range in CN strength is relatively small at a given
luminosity on the giant branch, in keeping with behavior seen in globular
clusters of [Fe/H] $< -1.8$. There is evidence of correlations between
CN and CH band strength among several giants.  Red giants in our LRISb sample
with [Fe/H] $> -2.25$ evince a trend of decreasing [C/Fe] with increasing 
luminosity, such as to afford evidence of the type of deep interior mixing that
is commonly found among both cluster and field metal-poor giants of the 
Galactic halo. The one subgroup of stars in our sample that possibly shows
disparate behavior with regard to CH and CH are those giants in the metallicity
range ${\rm [Fe/H]} < -2.25$, among which [C/Fe] appears to be greater
at higher luminosities. Observations of additional stars in this metallicity
range are needed to determine whether this is a consequence of small sample
statistics or is a general property of Draco giants of such metallicity.
Our LRISb Draco sample includes one star that potentially could be a 
CEMP star, however, we suggest that a follow-up spectrum be obtained to
determine if it truly belongs to the CEMP category and if there are any 
{\it r}- or {\it s}- process enhancements.

\acknowledgments

GHS acknowledges support from NSF grant AST-0908757.
We thank the staff of Keck Observatory for their assistance in obtaining the
observations reported in this paper, and the LRIS instrument team.
The authors wish to recognize and acknowledge the very significant cultural 
role and reverence that the summit of Mauna Kea has always had within the 
indigenous Hawaiian community.  We are most fortunate to have the opportunity 
to conduct observations from this mountain.

\clearpage

\begin{figure}
\epsscale{1.0}
\plotone{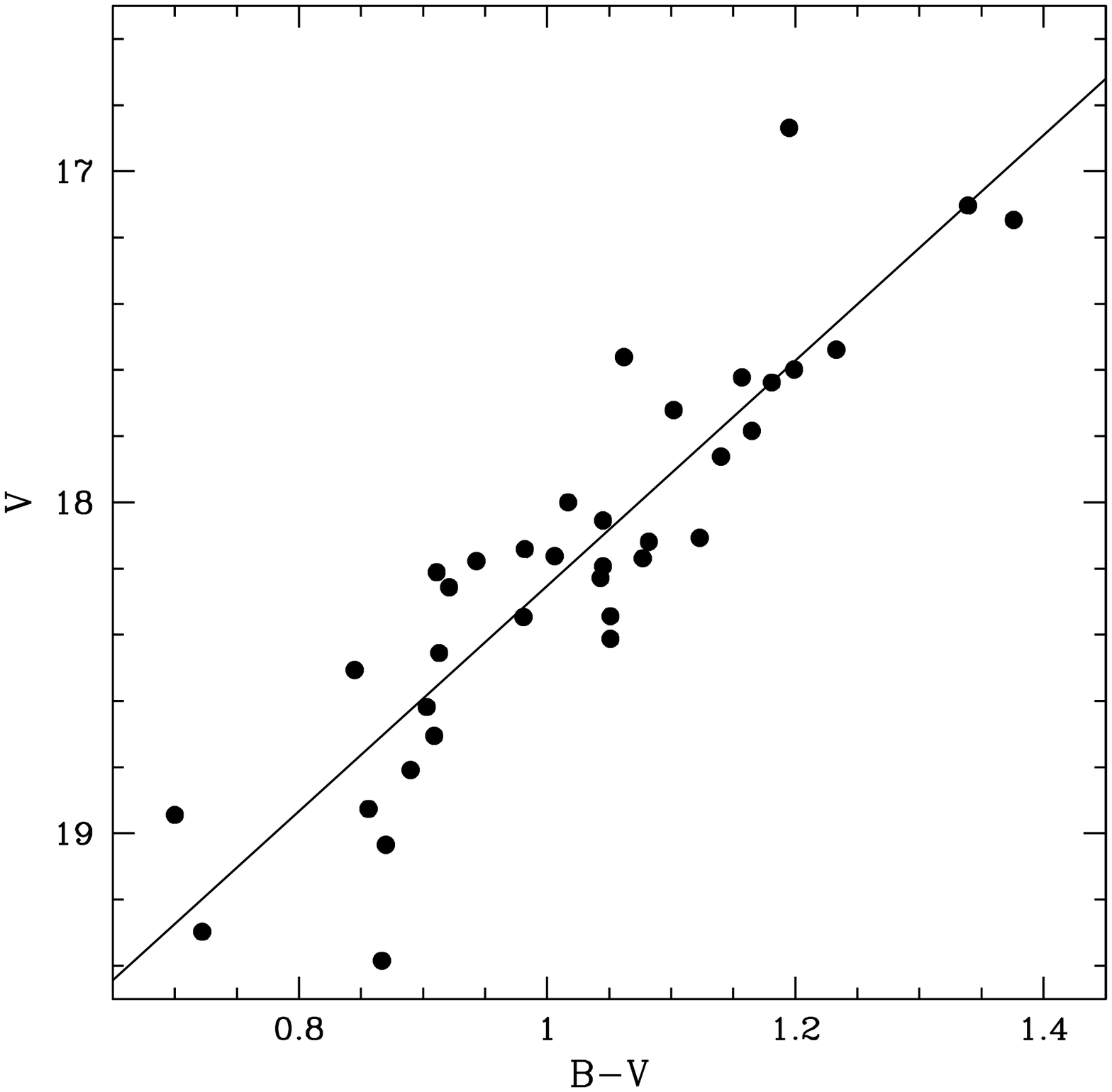}
\caption{The color-magnitude diagram of red giants observed in the Draco
dwarf spheroidal. A least-squares fit is shown by the solid line.}
\end{figure}
\clearpage

\begin{figure}
\begin{center}
\includegraphics[angle=-90,scale=0.6]{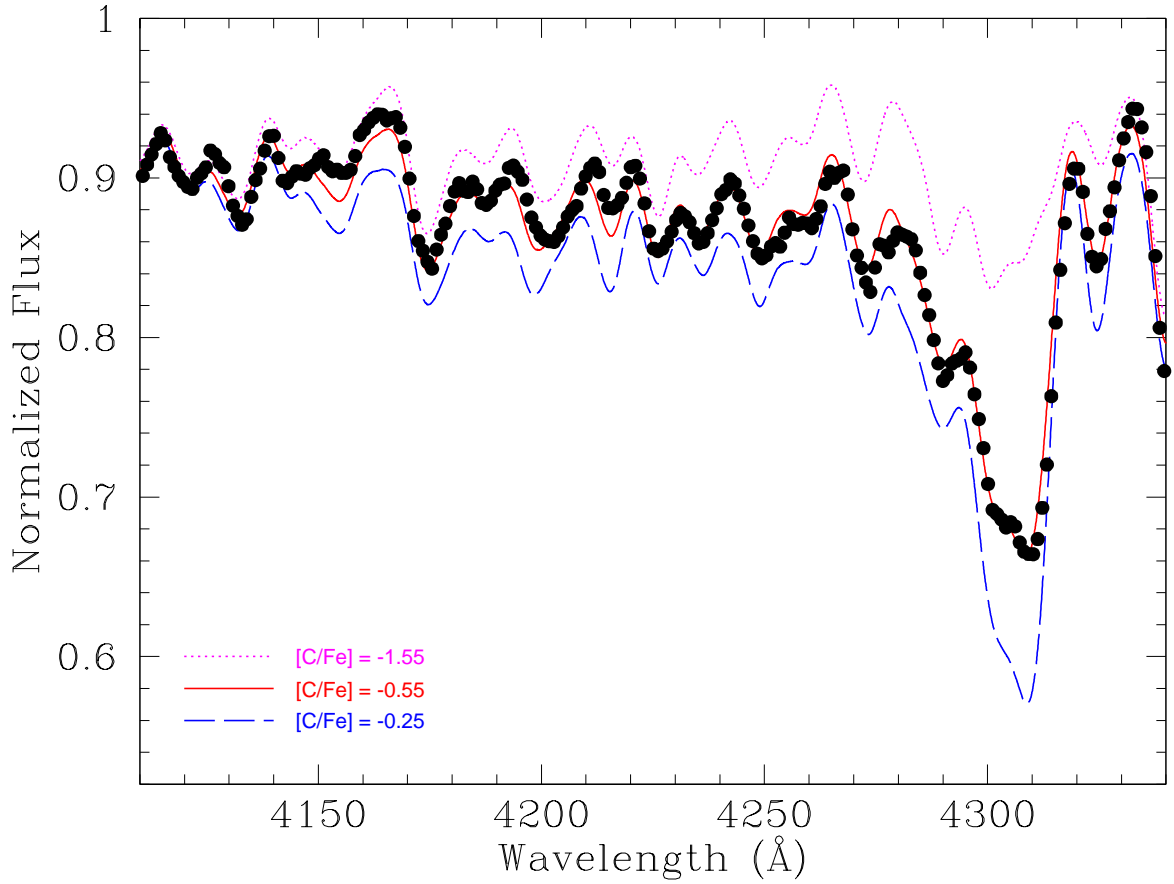}
\caption{The normalized spectrum for star 161 (star 3157 in Cohen \& Huang 
2009). The black dots represent the observed spectra and the three lines are 
synthetic spectra with different [C/Fe] ratios.}
\end{center}
\end{figure}
\clearpage

\begin{figure}
\epsscale{1.0}
\plotone{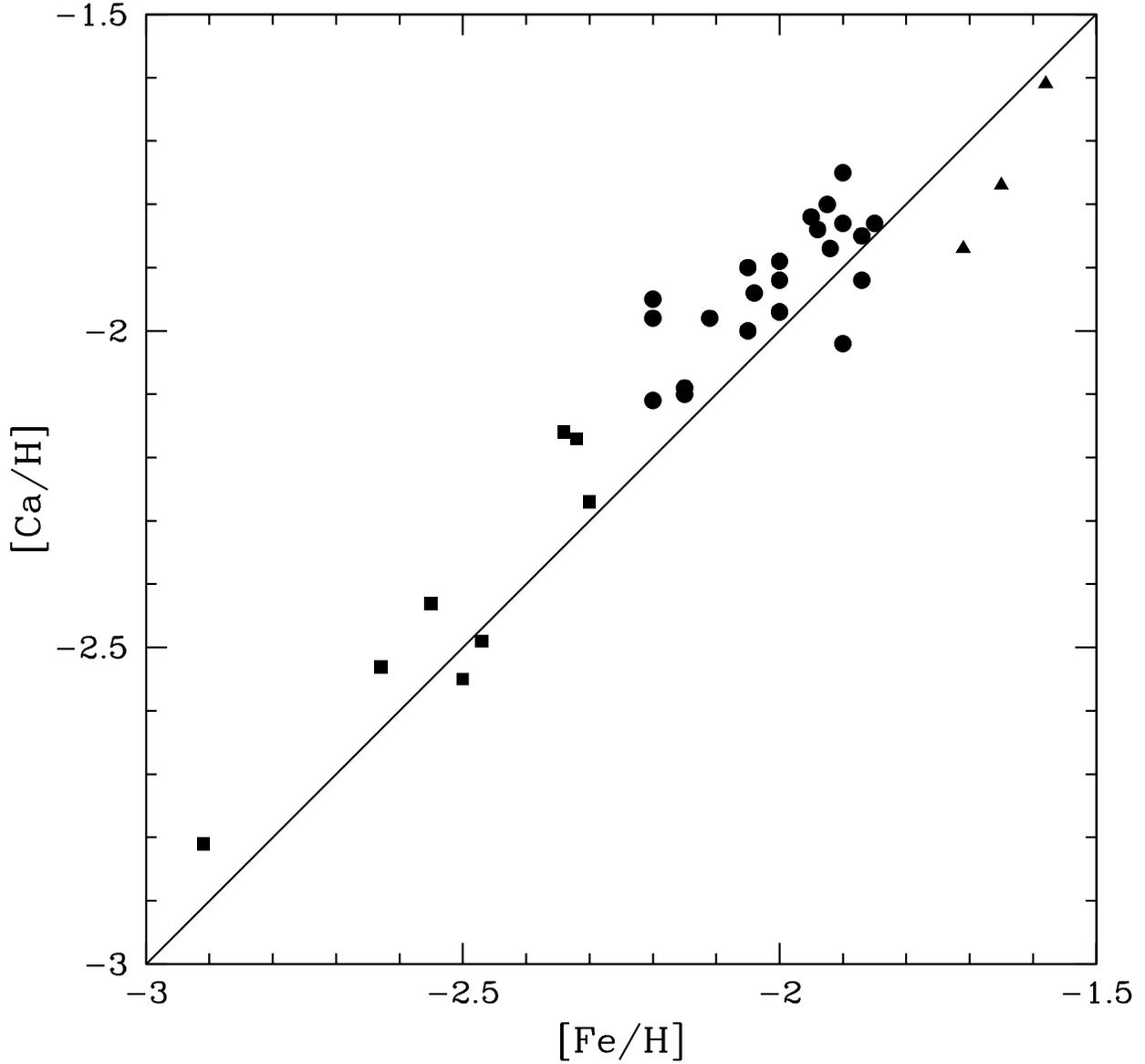}
\caption{Calcium verses iron abundance as derived from LRISb spectra.
The [Ca/H] and [Fe/H] values are from Columns 4 and 5 of Table 2.
Symbols are coded according to the [Fe/H] abundance:
filled squares (${\rm [Fe/H]} < -2.25$), 
filled circles ($-2.25 \leq {\rm [Fe/H]} \leq -1.85$), 
and filled triangles (${\rm [Fe/H]} > -1.85$). The solid line shows the locus 
for [Ca/H]=[Fe/H].}
\end{figure}
\clearpage

\begin{figure}
\epsscale{1.0}
\plotone{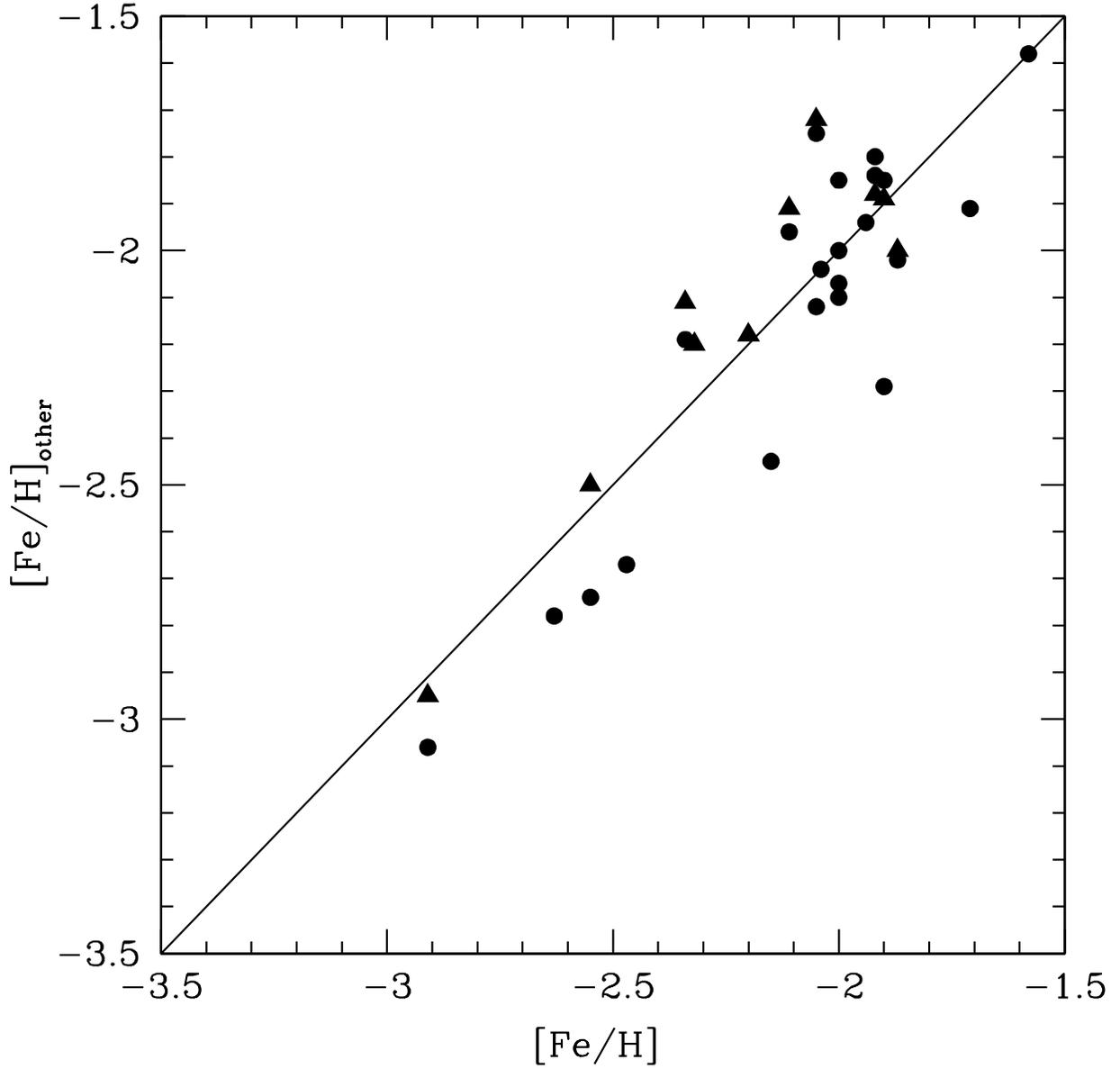}
\caption{Metallicities ${\rm [Fe/H]}_{\rm CaT}$ from the Ca triplet 
analysis of Winnick (2003) (red circles) and ${\rm [Fe/H]}$ 
from Kirby et al. (2010) (blue triangles) vs. [Fe/H] values derived 
from LRISb spectra (as listed in Table 2).}
\end{figure}
\clearpage

\begin{figure}
\epsscale{1.0}
\plotone{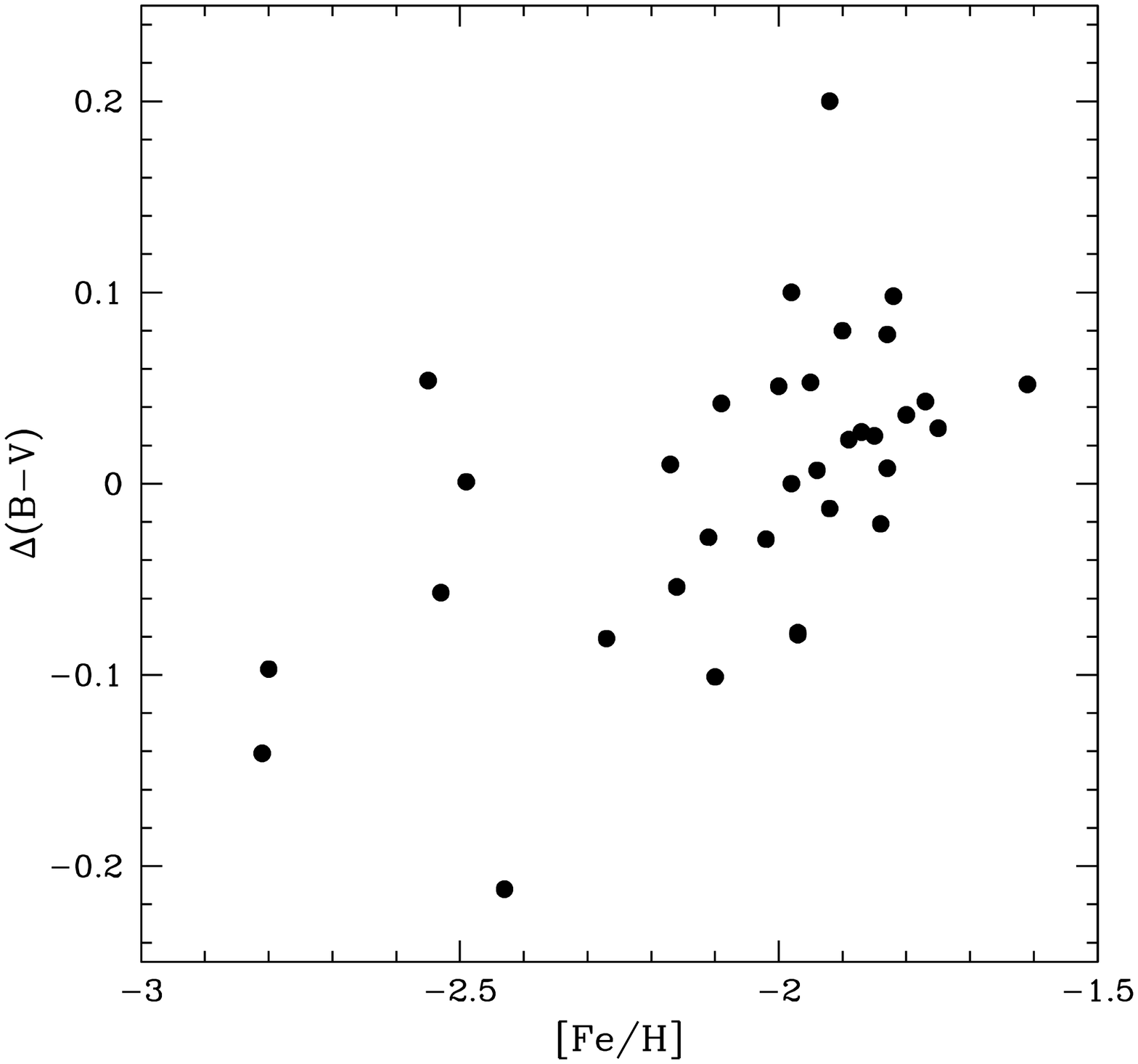}
\caption{The red giant branch color residual $\Delta (B-V)$ for Draco
stars in Table 2 vs. the metallicity [Fe/H] derived from LRISb spectra.}
\end{figure}
\clearpage

\begin{figure}
\epsscale{1.0}
\plotone{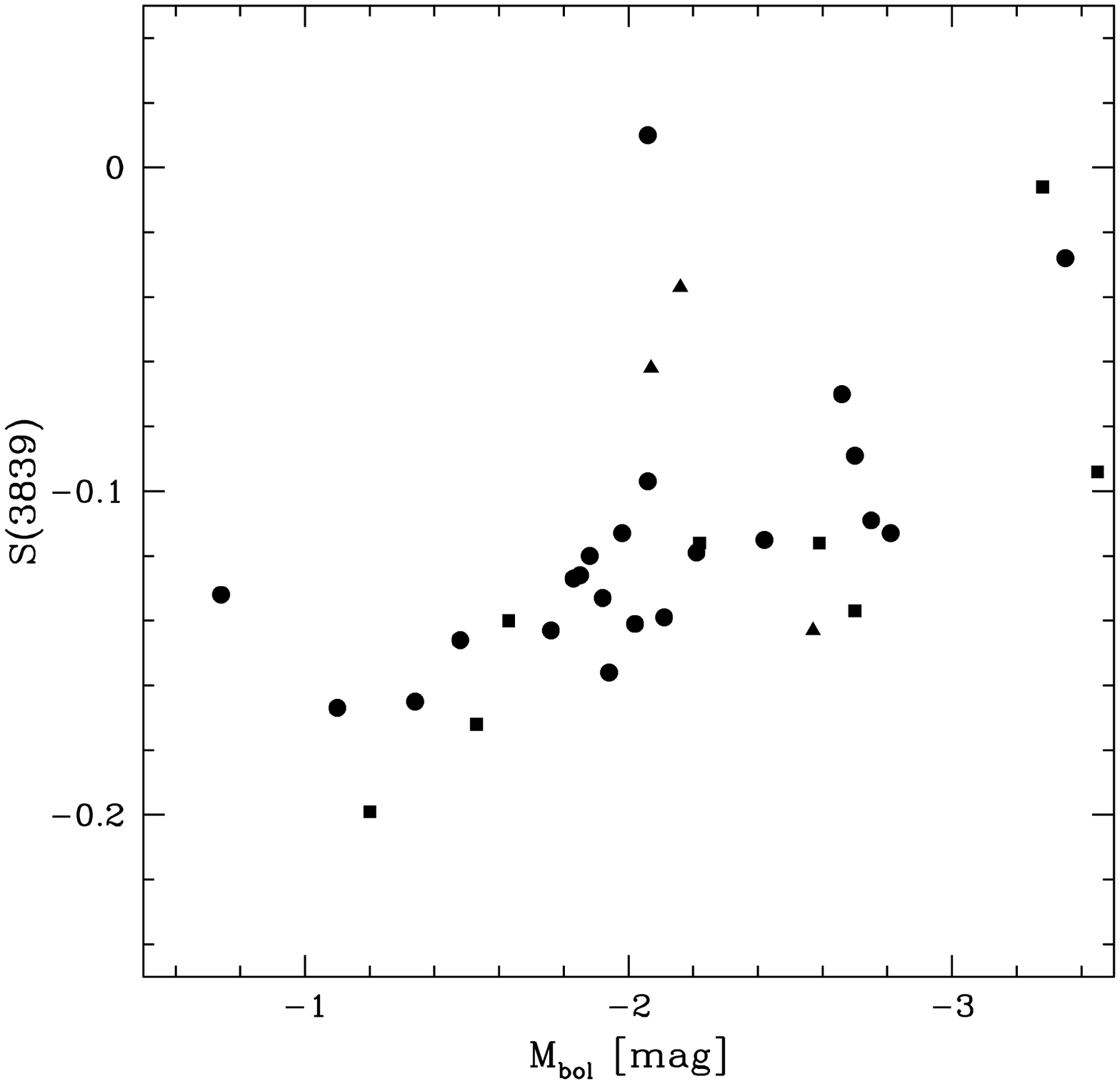}
\caption{The CN index $S(3839)$ vs. bolometric magnitude for the LRISb
sample of Draco giants. Symbols are used to denote [Fe/H] abundance
according to the convention of Figure 3,
i.e., filled squares (${\rm [Fe/H]} < -2.25$), 
filled circles ($-2.25 \leq {\rm [Fe/H]} \leq -1.85$), 
filled triangles (${\rm [Fe/H]} > -1.85$).
}
\end{figure}
\clearpage

\begin{figure}
\epsscale{1.0}
\plotone{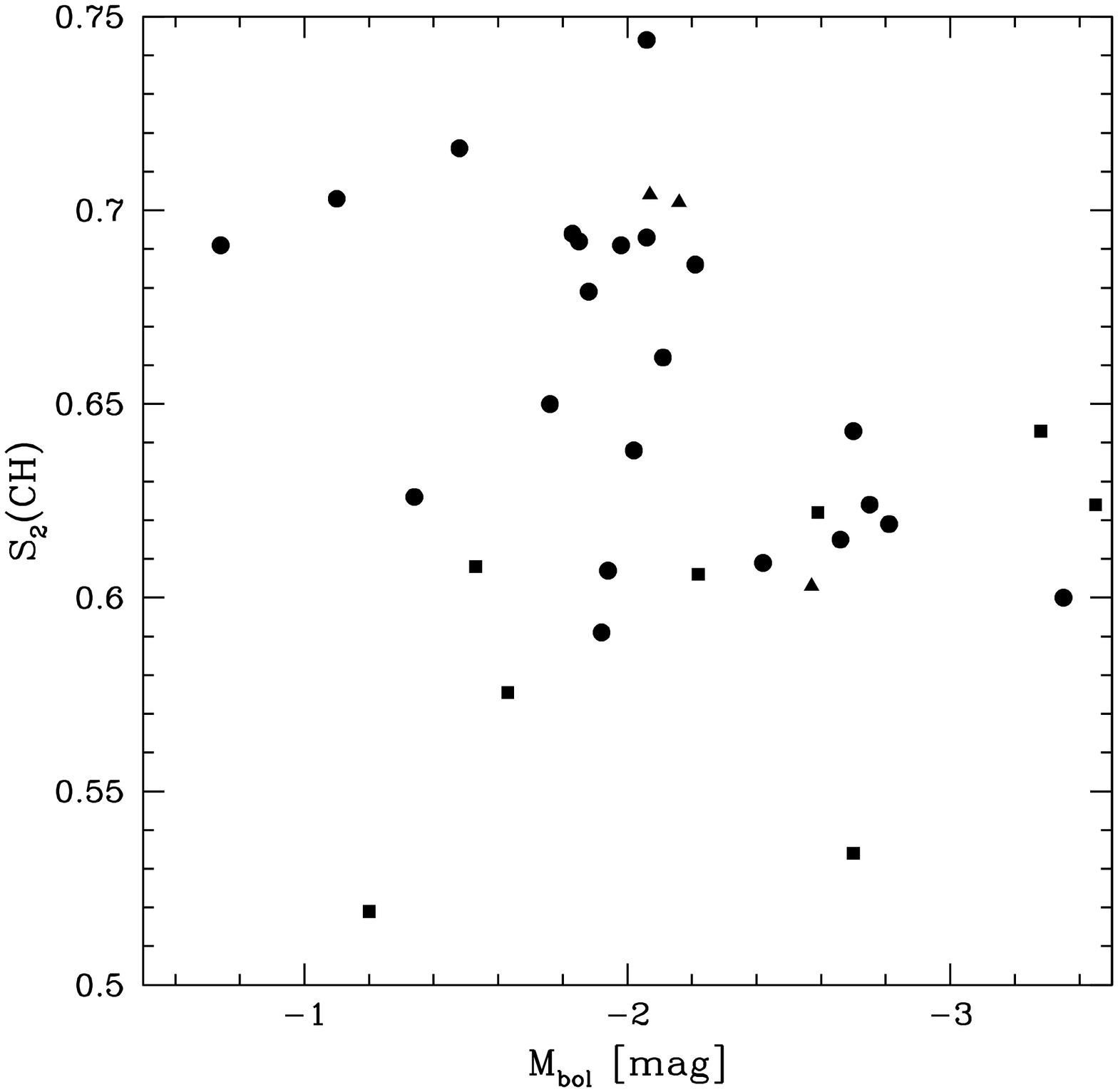}
\caption{The CH index $S_2$(CH) vs. bolometric magnitude for the LRISb
sample of Draco giants. Symbols denote [Fe/H] abundance according to the 
usage of Figure 3,
i.e., filled squares (${\rm [Fe/H]} < -2.25$), 
filled circles ($-2.25 \leq {\rm [Fe/H]} \leq -1.85$), 
filled triangles (${\rm [Fe/H]} > -1.85$).}
\end{figure}
\clearpage

\begin{figure}
\epsscale{1.0}
\plotone{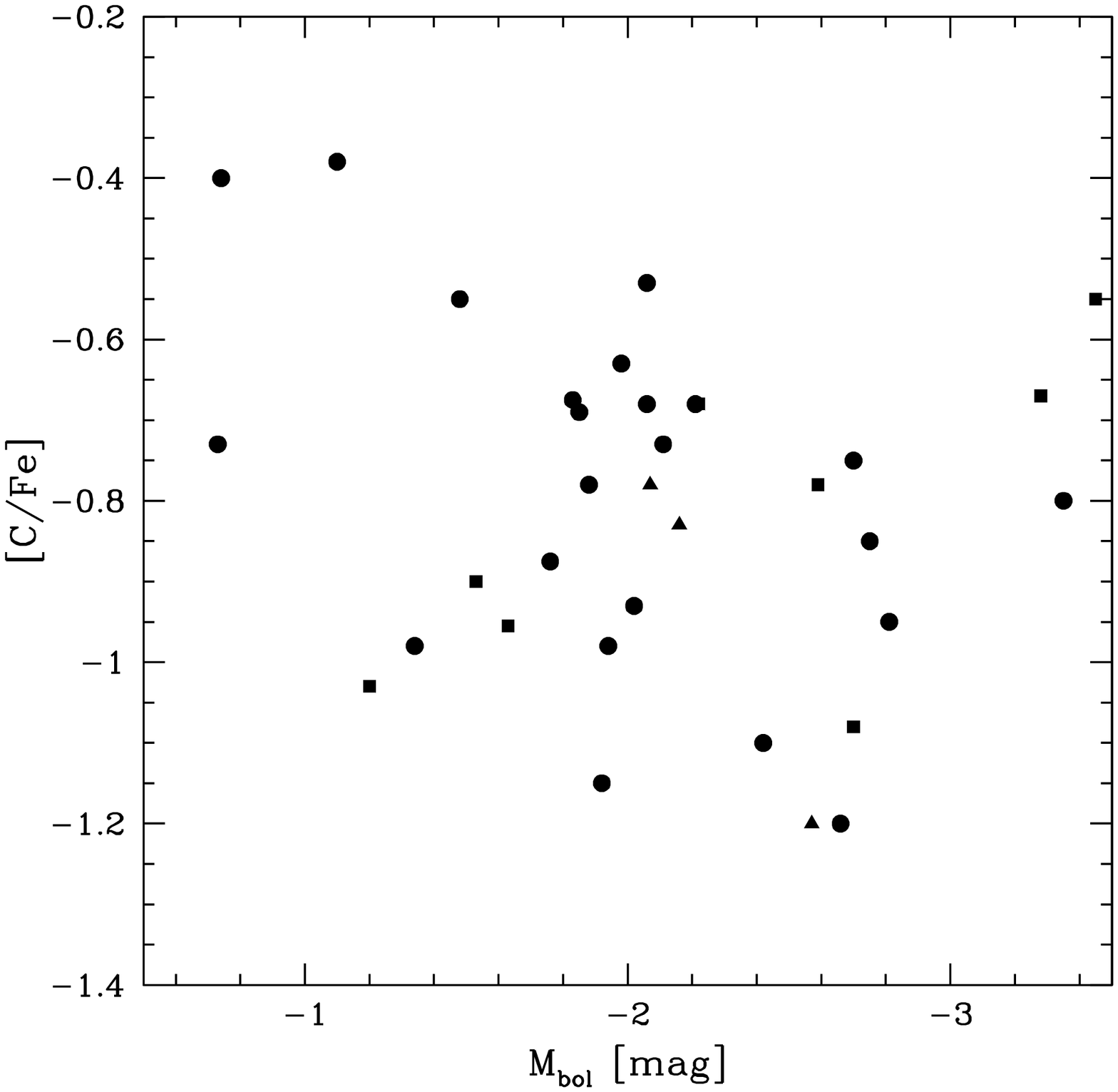}
\caption{Carbon abundance [C/Fe] vs. bolometric magnitude for giants in
the LRISb program. Symbols denote [Fe/H] abundance as in previous figures, 
i.e., filled squares (${\rm [Fe/H]} < -2.25$), 
filled circles ($-2.25 \leq {\rm [Fe/H]} \leq -1.85$), 
filled triangles (${\rm [Fe/H]} > -1.85$). Note that Star 589 
is not shown and would be off the top of the scale with [C/Fe]$=0.6$.}
\end{figure}
\clearpage

\begin{figure}
\epsscale{1.0}
\plotone{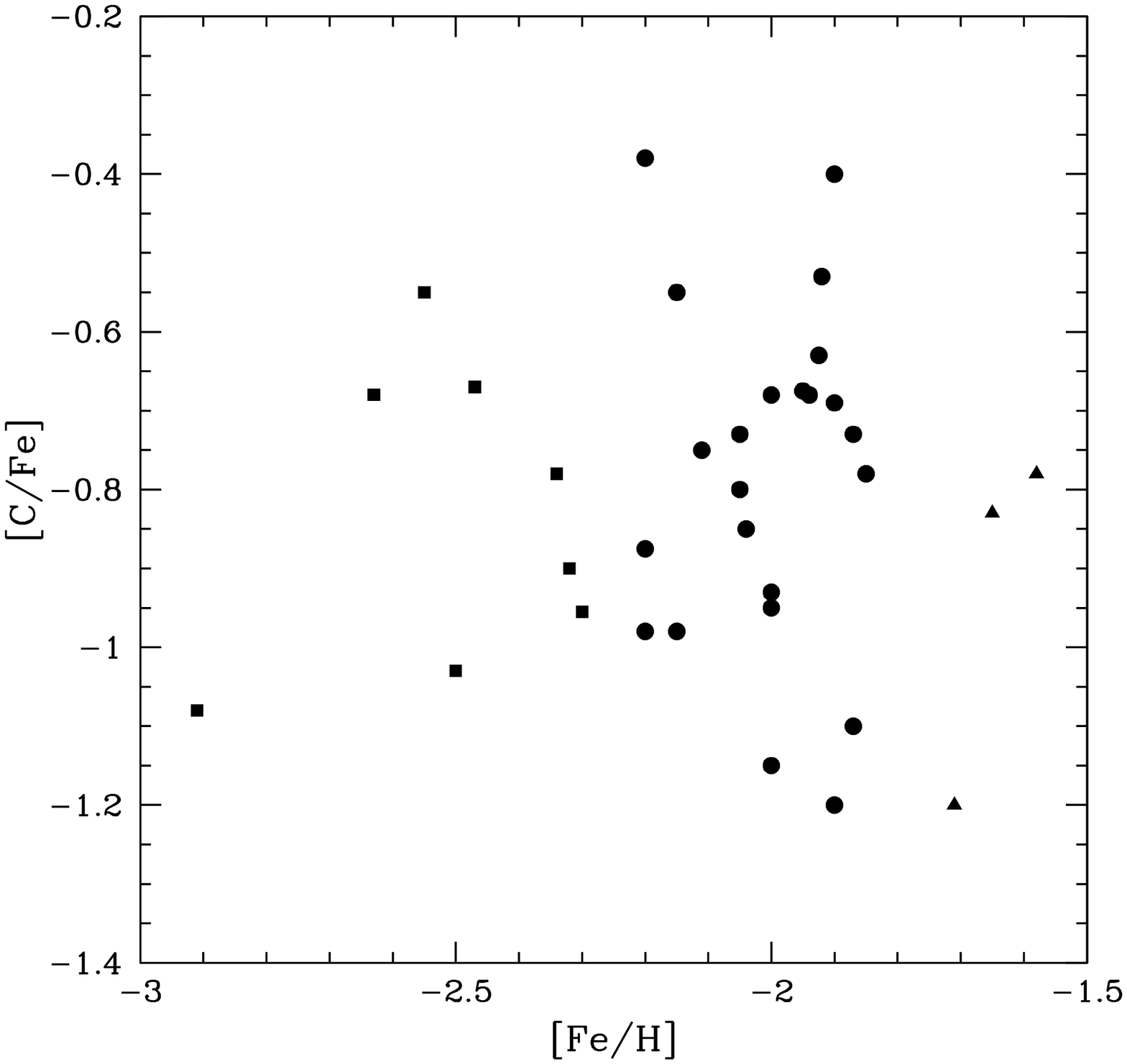}
\caption{Carbon abundance [C/Fe] vs. [Fe/H] for giants in the LRISb 
program. Symbols denote [Fe/H] abundance as in preceding figures. 
Note that Star 589 
is not shown and would be off the top of the scale with [C/Fe]$=0.6$.}
\end{figure}
\clearpage

\begin{figure}
\epsscale{1.0}
\plotone{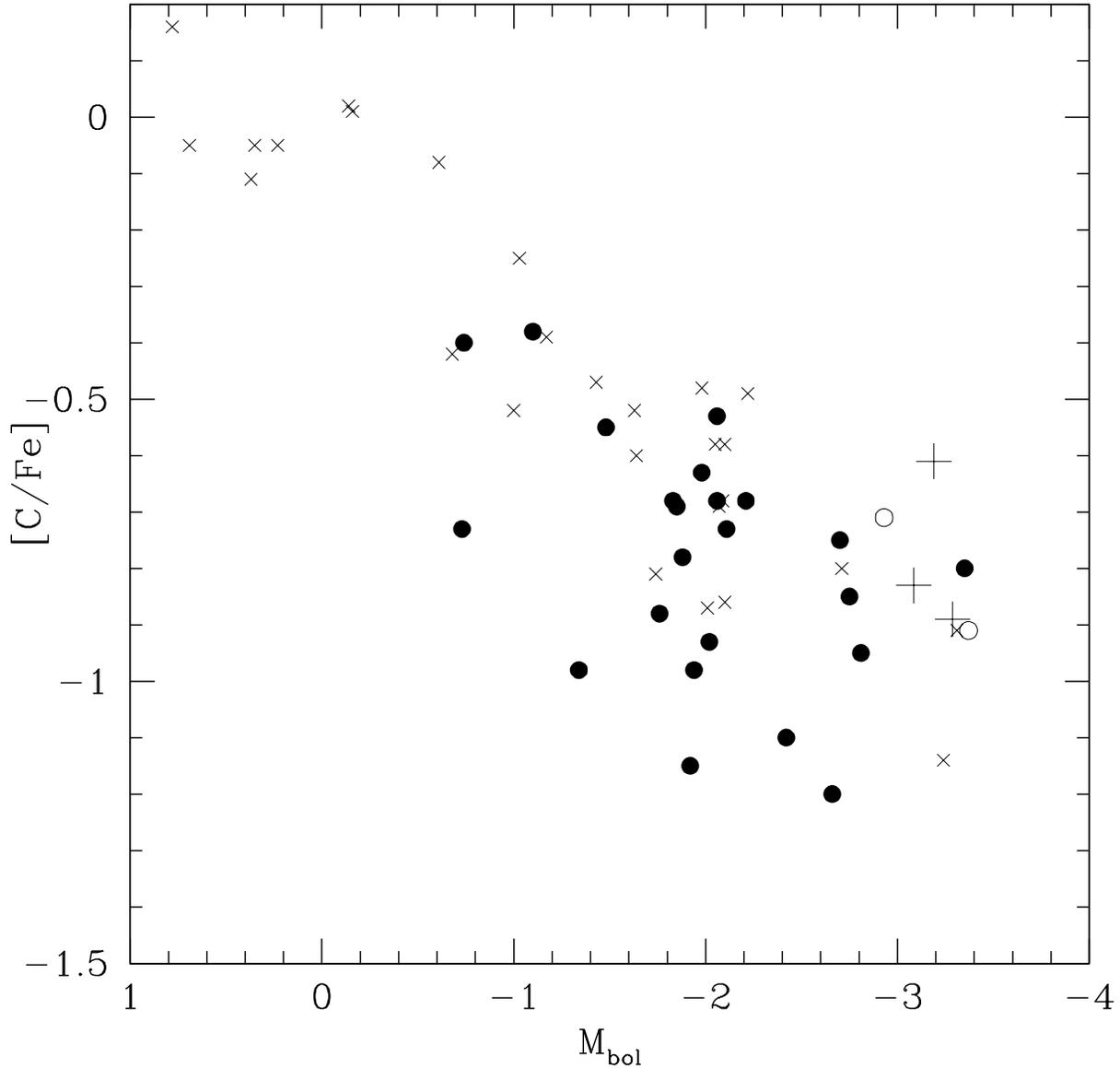}
\caption{Carbon abundance [C/Fe] vs. absolute bolometric magnitude for 
Draco giants (filled circles) with metallicities in the range 
$-2.25 \leq {\rm [Fe/H]} \leq -1.85$. Stars in the globular cluster
NGC 5466 are shown as a $\times$ symbol, with the data being taken from
Shetrone et al. (2010). The open circles are Draco stars also in the 
metallicity range $-2.25 \leq {\rm [Fe/H]} \leq -1.85$ taken from Cohen \& 
Huang (2009). The large crosses are Ursa Minor stars from Cohen \& Huang 
(2010) in the same metallicity range.}
\end{figure}
\clearpage

\begin{figure}
\epsscale{1.0}
\plotone{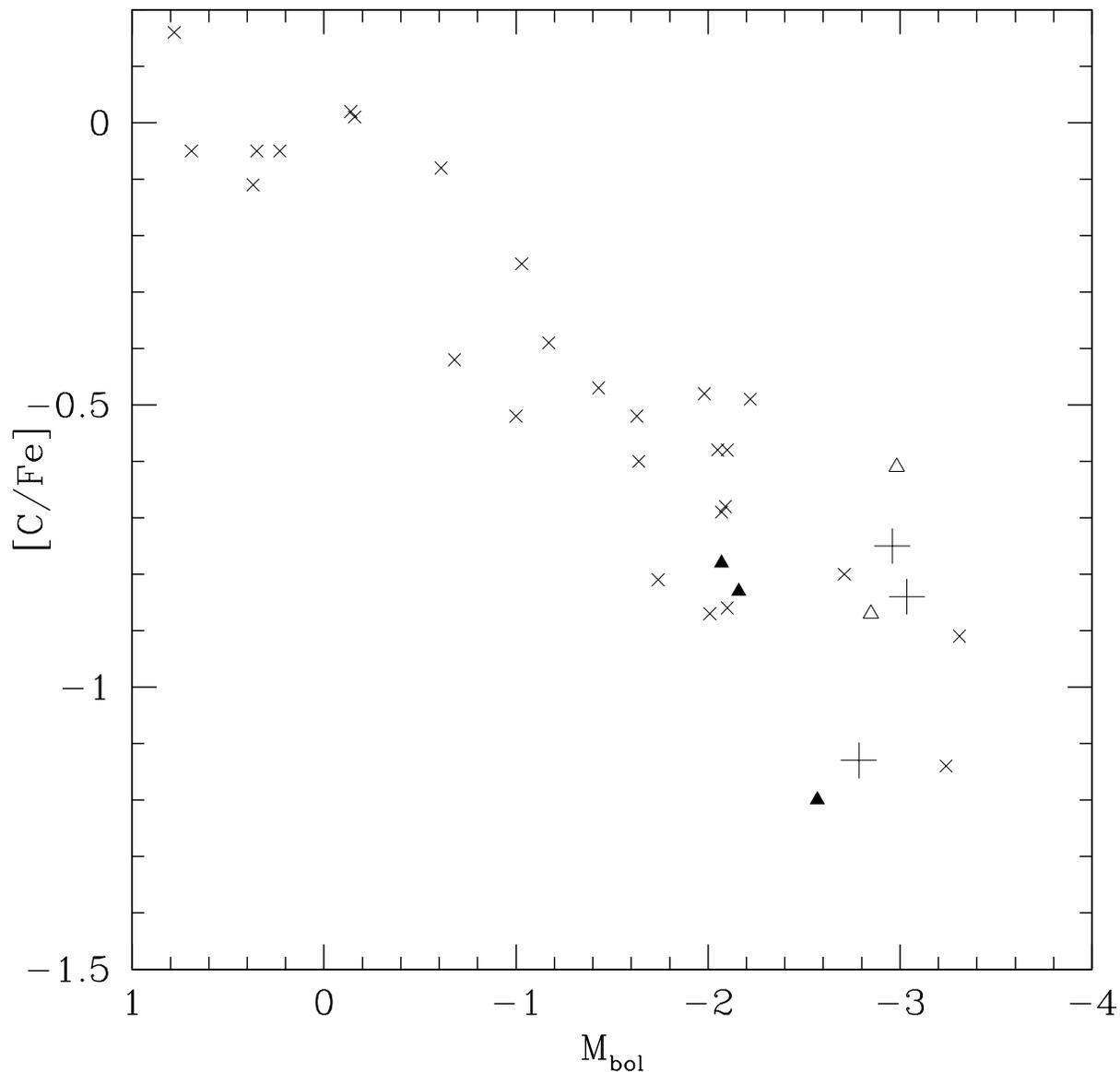}
\caption{Carbon abundance [C/Fe] vs. absolute bolometric magnitude for Draco
giants (filled triangles) with metallicities of ${\rm [Fe/H]} > -1.85$. The 
open triangles and large crosses are Draco and Ursa Minor stars respectively
taken from Cohen \& Huang (2009, 2010) in the same metallicity range. Stars 
in the globular cluster NGC 5466 (although of a lower metallicity than the
Draco stars plotted here) are shown as a $\times$ symbol, with the data being 
gleaned from Shetrone et al. (2010).}
\end{figure}
\clearpage

\begin{deluxetable}{rrccccccccc}
\tabletypesize{\scriptsize}
\tablecolumns{13}
\tablewidth{0pt}
\tablenum{1}
\tablecaption{Data for Giants Observed in the Draco Dwarf Spheroidal   \label{tbl1}}
\tablehead{
\colhead{ID}        &
\colhead{Alt ID}    &
\colhead{RA(2000)}  &
\colhead{Dec(2000)} &
\colhead{$V$}       &
\colhead{$(B-V)$}   &
\colhead{$(V-I)$}   &
\colhead{$J$}       & 
\colhead{$H$}       &
\colhead{$K$}       &
\colhead{${\rm [Fe/H]}_{\rm CaT}$\tablenotemark{1}} 
}
\startdata
     161 &     3157 & 17:19:41.85 & 57:52:19.5 & 16.87 &  1.20 &   1.29 &     14.70 &     14.12 &     14.09 &  --2.74    \\ 
     183 &       24 & 17:19:58.90 & 57:57:21.2 & 17.10 &  1.34 &   1.33 &     14.88 &     14.16 &     14.15 &  --2.67    \\ 
     187 &      267 & 17:19:44.75 & 57:57:37.3 & 17.15 &  1.38 &   1.44 &     14.74 &     13.92 &     13.89 &  --1.75    \\ 
     235 &      119 & 17:20:16.14 & 57:52:56.3 & 17.56 &  1.06 &   1.22 &     15.56 &     15.12 &     14.85 &  --3.06    \\ 
     237 &      490 & 17:19:39.97 & 57:54:25.1 & 17.60 &  1.20 &   1.24 &     15.42 &     14.74 &     14.65 &  --2.04    \\ 
     239 &      449 & 17:19:51.82 & 57:59:18.0 & 17.54 &  1.23 &   1.25 &     15.31 &     14.85 &     14.51 &  --2.00    \\ 
     240 &       11 & 17:20:05.68 & 57:57:53.0 & 17.64 &  1.18 &   1.27 &     15.51 &     14.81 &     14.68 &  --1.96    \\ 
     249 &    22209 & 17:20:21.13 & 57:49:27.5 & 17.62 &  1.16 &   1.31 &     15.58 &     15.02 &     14.96 &  --2.29    \\ 
     262 &       45 & 17:19:57.92 & 57:56:58.6 & 17.72 &  1.10 &   1.24 &     15.69 &     14.83 &     14.81 &  --2.19    \\ 
     276 &      286 & 17:19:45.14 & 57:55:14.5 & 17.78 &  1.17 &   1.29 &     15.66 &     14.90 &     14.73 &  --1.91    \\ 
     285 &      297 & 17:19:41.16 & 57:54:56.9 & 17.86 &  1.14 &   1.18 &     15.81 &     15.14 &     15.12 &  --2.02    \\ 
     314 &      506 & 17:19:53.05 & 57:51:38.1 & 18.00 &  1.02 &   1.18 &     16.30 &     15.43 &     15.49 &  --2.78    \\ 
     325 &  \nodata & 17:20:16.99 & 57:53:12.5 & 18.12 &  1.08 &   1.24 &     16.18 &     15.43 &     15.28 &   \nodata \\ 
     326 &      281 & 17:19:43.49 & 57:56:33.5 & 18.05 &  1.04 &   1.17 &     16.02 &     15.44 &     15.18 &  --1.85    \\ 
     327 &      522 & 17:20:13.39 & 57:50:51.9 & 18.11 &  1.12 &   1.15 &     16.31 &     15.68 &     15.42 &  --2.12    \\ 
     330 &     3213 & 17:20:11.64 & 57:49:36.6 & 18.14 &  0.98 &   1.20 &     16.16 &     15.59 &     15.43 &  --2.35    \\ 
     334 &      462 & 17:19:43.01 & 57:58:39.7 & 18.16 &  1.01 &   1.20 &     16.14 &     15.71 &     15.66 &  --1.94    \\ 
     337 &     3210 & 17:20:05.33 & 57:50:18.4 & 18.18 &  0.94 &   1.14 &     16.25 &     15.75 &     15.59 &  --2.10    \\ 
     348 &      335 & 17:20:06.92 & 57:52:37.4 & 18.19 &  1.04 &   1.20 &     15.94 &     15.84 &     15.47 &  --1.84    \\ 
     354 &       22 & 17:20:01.60 & 57:57:04.8 & 18.17 &  1.08 &   1.15 &     16.25 &     15.65 &     15.35 &  --1.58    \\ 
     361 &        K & 17:19:55.79 & 57:53:48.9 & 18.23 &  1.04 &   1.18 &     16.40 &     15.69 &     15.73 &  --1.80    \\ 
     363 &        H & 17:20:15.72 & 57:53:43.5 & 18.21 &  0.91 &   1.08 &     16.54 &     15.93 &     15.67 &  --2.45    \\ 
     368 &      350 & 17:20:20.39 & 57:51:58.5 & 18.26 &  0.92 &   1.10 &     16.45 &     15.77 &   \nodata &  --2.07    \\ 
     386 &      273 & 17:19:50.05 & 57:56:40.9 & 18.35 &  0.98 &   1.16 &     16.54 &     16.04 &     15.56 &  --1.85    \\ 
     389 &  \nodata & 17:19:53.46 & 57:56:16.7 & 18.34 &  1.05 &   1.10 &     16.55 &     15.52 &     15.66 &   \nodata \\ 
     409 &  \nodata & 17:19:56.60 & 57:52:43.1 & 18.45 &  0.91 &   1.13 &     16.50 &     15.80 &   \nodata &   \nodata \\ 
     410 &  \nodata & 17:19:57.66 & 57:54:35.4 & 18.41 &  1.05 &   1.11 &     16.40 &     15.78 &   \nodata &   \nodata \\ 
     427 &  \nodata & 17:19:56.92 & 57:52:23.2 & 18.51 &  0.84 &   1.06 &     16.70 &     16.17 &   \nodata &   \nodata \\ 
     482 &  \nodata & 17:20:24.98 & 57:54:50.9 & 18.70 &  0.91 &   1.12 &   \nodata &   \nodata &   \nodata &   \nodata \\ 
     483 &  \nodata & 17:19:57.03 & 57:52:58.1 & 18.62 &  0.90 &   1.00 &   \nodata &   \nodata &   \nodata &   \nodata \\ 
     546 &  \nodata & 17:20:01.96 & 57:51:30.6 & 18.81 &  0.89 &   1.03 &   \nodata &   \nodata &   \nodata &   \nodata \\ 
     589 &  \nodata & 17:19:50.23 & 57:53:15.5 & 18.94 &  0.70 &   1.02 &   \nodata &   \nodata &   \nodata &   \nodata \\ 
     621 &  \nodata & 17:20:15.42 & 57:53:30.5 & 18.93 &  0.86 &   0.93 &   \nodata &   \nodata &   \nodata &   \nodata \\ 
     643 &  \nodata & 17:20:24.11 & 57:55:15.7 & 19.03 &  0.87 &   1.01 &   \nodata &   \nodata &   \nodata &   \nodata \\ 
     775 &  \nodata & 17:20:02.77 & 57:48:57.4 & 19.30 &  0.72 &   0.93 &   \nodata &   \nodata &   \nodata &   \nodata \\ 
     810 &  \nodata & 17:19:50.28 & 57:55:20.4 & 19.39 &  0.87 &   1.00 &   \nodata &   \nodata &   \nodata &   \nodata \\ 
\enddata 
\tablenotetext{1}{Winnick (2003)}
\end{deluxetable} 

\clearpage

\begin{deluxetable}{lrrccccccccrcc}
\tabletypesize{\scriptsize}
\tablecolumns{14}
\tablewidth{0pt}
\tablenum{2}
\tablecaption{Indices and Abundances for Draco dSph Red Giants} 
\label{tbl2}
\tablehead{
\colhead{Star}        &
\colhead{$S(3839)$}   &
\colhead{$S_2$(CH)}   &
\colhead{[Fe/H]}      &
\colhead{$\epsilon$[Fe/H]}  &
\colhead{[Ca/H]}      &
\colhead{$\epsilon$[Ca/H]}  &
\colhead{[C/Fe]}      &
\colhead{$\epsilon$[C/Fe]}  &
\colhead{$M_{\rm bol}$}  &
\colhead{$T_{\rm eff}$}  &
\colhead{$\epsilon (T_{\rm eff})$\tablenotemark{a}}  &
\colhead{$\log g$}    &
\colhead{$v_t$}       \\
\colhead{1}           &
\colhead{2}           &
\colhead{3}           & 
\colhead{4}           & 
\colhead{5}           & 
\colhead{6}           & 
\colhead{7}           & 
\colhead{8}           & 
\colhead{9}           & 
\colhead{10}          & 
\colhead{11}          &
\colhead{12}          &
\colhead{13}          &
\colhead{14}          
}
\startdata
161   & --0.094 & 0.624 & --2.55 &0.15& --2.43 &0.09& --0.55 &0.21& --3.45 & 4378 &  85 & 0.52 & 1.94 \\ 
183   & --0.006 & 0.643 & --2.47 &0.16& --2.49 &0.09& --0.67 &0.22& --3.28 & 4279 &  93 & 0.55 & 1.93 \\ 
187   & --0.028 & 0.600 & --2.05 &0.15& --2.00 &0.09& --0.80 &0.21& --3.35 & 4127 &  82 & 0.46 & 1.96 \\
235   & --0.137 & 0.534 & --2.91 &0.15& --2.81 &0.09& --1.08 &0.21& --2.70 & 4533 &  85 & 0.88 & 1.79 \\
237   & --0.109 & 0.624 & --2.04 &0.15& --1.94 &0.09& --0.85 &0.21& --2.75 & 4327 &  84 & 0.78 & 1.83 \\
239   & --0.113 & 0.619 & --2.00 &0.16& --1.89 &0.09& --0.95 &0.21& --2.81 & 4324 &  87 & 0.75 & 1.84 \\
240   & --0.089 & 0.643 & --2.11 &0.15& --1.98 &0.09& --0.75 &0.21& --2.70 & 4334 &  82 & 0.80 & 1.82 \\
249   & --0.070 & 0.615 & --1.90 &0.16& --2.02 &0.09& --1.20 &0.22& --2.66 & 4436 &  90 & 0.86 & 1.80 \\
262   & --0.116 & 0.622 & --2.34 &0.16& --2.16 &0.09& --0.78 &0.21& --2.59 & 4383 &  88 & 0.87 & 1.80 \\ 
276   & --0.143 & 0.603 & --1.71 &0.15& --1.87 &0.09& --1.20 &0.21& --2.57 & 4312 &  83 & 0.85 & 1.80 \\
285   & --0.115 & 0.609 & --1.87 &0.15& --1.85 &0.09& --1.10 &0.21& --2.42 & 4428 &  85 & 0.95 & 1.76 \\
314   & --0.116 & 0.606 & --2.63 &0.18& --2.53 &0.11& --0.68 &0.24& --2.22 & 4611 & 110 & 1.10 & 1.70 \\
325   & --0.037 & 0.702 & --1.65 &0.15& --1.77 &0.09& --0.83 &0.21& --2.16 & 4435 &  86 & 1.06 & 1.72 \\
326   & --0.119 & 0.686 & --2.00 &0.15& --1.92 &0.09& --0.68 &0.21& --2.21 & 4453 &  83 & 1.04 & 1.72 \\
327   & --0.139 & 0.662 & --2.05 &0.17& --1.90 &0.10& --0.73 &0.23& --2.11 & 4558 & 102 & 1.13 & 1.69 \\
334   & --0.097 & 0.693 & --1.94 &0.16& --1.84 &0.09& --0.68 &0.21& --2.06 & 4545 &  88 & 1.14 & 1.68 \\
337   & --0.141 & 0.638 & --2.00 &0.15& --1.97 &0.09& --0.93 &0.21& --2.02 & 4597 &  81 & 1.18 & 1.67 \\
348   &   0.010 & 0.744 & --1.92 &0.18& --1.87 &0.10& --0.53 &0.23& --2.06 & 4471 & 104 & 1.11 & 1.69 \\
354   & --0.062 & 0.704 & --1.58 &0.16& --1.61 &0.09& --0.78 &0.21& --2.07 & 4500 &  87 & 1.12 & 1.69 \\
361\tablenotemark{b} & --0.113 & 0.691 &--1.92 &0.16& --1.80 &0.09& --0.63 & 0.22 & --1.98 & 4567 &  92 & 1.18 & 1.67 \\
363   & --0.156 & 0.607 & --2.15 &0.17& --2.10 &0.10 &--0.98 &0.22 &--1.94 & 4730 &  97 & 1.26 & 1.63 \\
368   & --0.133 & 0.591 & --2.00 &0.16& --1.97 &0.09 &--1.15 &0.21 &--1.92 & 4642 &  87 & 1.23 & 1.64 \\
386   & --0.126 & 0.692 & --1.90 &0.17& --1.83 &0.10 &--0.69 &0.23 &--1.85 & 4597 &  99 & 1.25 & 1.64 \\
389   & --0.120 & 0.679 & --1.85 &0.18& --1.83 &0.10 &--0.78 &0.23 &--1.88 & 4527 & 108 & 1.20 & 1.66 \\
409\tablenotemark{b} & --0.143 & 0.650 & --2.20 &0.15& --2.11 &0.09& --0.87 &0.21& --1.76 & 4556 &  92 & 1.26 & 1.63 \\
410\tablenotemark{b} & --0.127 & 0.694 & --1.95 &0.16& --1.82 &0.09& --0.67 &0.21& --1.83 & 4498 &  87 & 1.22 & 1.65 \\
427\tablenotemark{b} & --0.140 & 0.576 & --2.30 &0.15& --2.27 &0.09& --0.95 &0.21& --1.63 & 4725 &  82 & 1.38 & 1.58 \\
482   & --0.146 & 0.716 & --2.15 &0.15& --2.09 &0.09& --0.55 &0.21& --1.48 & 4623 &  85 & 1.40 & 1.58 \\
483   & --0.172 & 0.608 & --2.32 &0.19& --2.17 &0.11& --0.90 &0.24& --1.53 & 4734 & 115 & 1.42 & 1.57 \\
546   & --0.165 & 0.626 & --2.20 &0.16& --1.95 &0.09& --0.98 &0.21& --1.34 & 4717 &  88 & 1.49 & 1.54 \\
589   &    .... &  .... & --2.80 &0.25&   .... &....&  +0.60 &0.30& --1.12 & 4920 & 166 & 1.65 & 1.47 \\
621   & --0.199 & 0.519 & --2.50 &0.23& --2.55 &0.13& --1.03 &0.28& --1.20 & 4826 & 148 & 1.59 & 1.50 \\
643   & --0.167 & 0.703 & --2.20 &0.16& --1.98 &0.09& --0.38 &0.21& --1.10 & 4756 &  87 & 1.60 & 1.49 \\
775   & --0.132 & 0.691 & --1.90 &0.15& --1.75 &0.09& --0.40 &0.21& --0.74 & 4987 &  86 & 1.83 & 1.40 \\
810   &   ....  &  .... & --1.87 &0.15& --1.92 &0.09& --0.73 &0.21& --0.73 & 4774 &  83 & 1.76 & 1.43 \\
\enddata
\tablenotetext{a}{These $T_{\rm eff}$ errors have been convolved with 80K; see text.} 
\tablenotetext{b}{Based on spectra from two slit masks.} 
\end{deluxetable}

\clearpage

\begin{deluxetable}{crccccccccc}
\tabletypesize{\small}
\tablecolumns{10}
\tablewidth{0pt}
\tablenum{3}
\tablecaption{Comparisons Between Abundances} 
\label{tbl3}
\tablehead{
\colhead{LRISb ID} &
\colhead{Alt ID}   &
\colhead{[Fe/H]}  &
\colhead{[C/Fe]}  &
\colhead{[Fe/H]}  &
\colhead{[Fe/H]}  &
\colhead{[Fe/H]}  &
\colhead{[C/Fe]}  &
\colhead{[Fe/H]}  &
\colhead{[C/Fe]}  &
\colhead{[Fe/H]}  \nl
\colhead{}        &
\colhead{}        &
\colhead{LRISb}   &
\colhead{LRISb}   &
\colhead{SCS\tablenotemark{a}}   &
\colhead{Winnick} &
\colhead{C\&H\tablenotemark{b}}  &
\colhead{C\&H\tablenotemark{b}}  &
\colhead{FRC\tablenotemark{c}}   &
\colhead{FRC\tablenotemark{c}}   &
\colhead{K et al.\tablenotemark{d}}
}
\startdata
161  &  3157  &  --2.55  &  --0.55  &   ....   &   ....   &  --2.45  &  --0.29  &     ....  &    ....   &  --2.50  \\
183  &    24  &  --2.47  &  --0.67  &  --2.36  &  --2.67  &    ....  &    ....  &     ....  &    ....   &    ....  \\   
187  &   267  &  --2.05  &  --0.80  &  --1.67  &  --1.75  &    ....  &    ....  &     ....  &    ....   &  --1.72   \\   
235  &   119  &  --2.91  &  --1.08  &  --2.97  &  --3.06  &    ....  &    ....  &   --2.95  &  --0.48   &  --2.95  \\  
240  &    11  &  --2.11  &  --0.95  &  --1.72  &  --1.96  &    ....  &    ....  &     ....  &    ....   &  --1.91  \\  
262  &    45  &  --2.34  &  --0.78  &    ....  &    ....  &    ....  &    ....  &     ....  &    ....   &  --2.11  \\
348  &   335  &  --1.92  &  --0.53  &    ....  &    ....  &    ....  &    ....  &     ....  &    ....   &  --1.88  \\
386  &   273  &  --1.90  &  --0.69  &    ....  &    ....  &    ....  &    ....  &     ....  &    ....   &  --1.89  \\
483  &  ....  &  --2.32  &  --0.90  &    ....  &    ....  &    ....  &    ....  &     ....  &    ....   &  --2.20  \\
546  &  ....  &  --2.20  &  --0.98  &    ....  &    ....  &    ....  &    ....  &     ....  &    ....   &  --2.18  \\
810  &  ....  &  --1.87  &  --0.73  &    ....  &    ....  &    ....  &    ....  &     ....  &    ....   &  --2.00  \\
\enddata
\tablenotetext{a}{SCS: Shetrone et al. (2001a)}
\tablenotetext{b}{C\&H: Cohen \& Huang (2009)}
\tablenotetext{c}{FRC: Fulbright et al. (2004)}
\tablenotetext{d}{K et al: Kirby et al. (2010)}
\end{deluxetable}

\end{document}